\documentclass[aps,prb,twocolumn,superscriptaddress,footinbib]{revtex4-2}
\usepackage[colorlinks=true,citecolor=black,urlcolor=black,linkcolor=black,pdfstartview=FitH,bookmarksopen]{hyperref}
\usepackage{graphicx}
\usepackage[nointegrals]{wasysym} 
\usepackage[export]{adjustbox}
\usepackage{amsmath,amsfonts,amssymb,latexsym}
\usepackage{hhline}
\usepackage{bm}
\usepackage{verbatim}
\usepackage{enumitem}
\usepackage{mathrsfs}
\usepackage{slashed}
\usepackage{empheq}

\newcommand{\pt}{{\partial}}

\newcommand{\xv}{{\bf x}}

\newcommand{\zh}{{\hat{\bf z}}}

\newcommand{\oh}{{\frac{1}{2}}}

\newcommand{\cH}{{\mathcal H}}
\newcommand{\cL}{{\mathcal L}}
\newcommand{\grad}{{\bm{\nabla}}}

\newcommand{\be}{\begin{equation}}
\newcommand{\ee}{\end{equation}}
\newcommand{\bea}{\begin{eqnarray}}
\newcommand{\eea}{\end{eqnarray}}
\newcommand{\bse}{\begin{subequations}}
\newcommand{\ese}{\end{subequations}}
\def\rf#1{(\ref{#1})}

\renewcommand{\(}{\left(}
\renewcommand{\)}{\right)}
\renewcommand{\[}{\left[}
\renewcommand{\]}{\right]}

\renewcommand\d\partial

\newcommand\bpm            {\begin{pmatrix}}
\newcommand\epm           {\end{pmatrix}}

\def\app#1#2{	\mathrel{		\setbox0=\hbox{$#1\sim$}		\setbox2=\hbox{			\rlap{\hbox{$#1\propto$}}			\lower1.1\ht0\box0		}		\raise0.25\ht2\box2	}}

\newcommand{\ope}\odot

\usepackage{manfnt}

\newcommand{\bi}{\begin{itemize}}
	\newcommand{\ei}{\end{itemize}}

\usepackage{amsthm}

\newtheorem{theorem}{Theorem}

\theoremstyle{definition}
\newtheorem{definition}{Definition}
\theoremstyle{definition}

\newcommand\bd            {\begin{definition}}
	\newcommand\ed            {\end{definition}}
\newcommand\bt            {\begin{theorem}}
	\newcommand\et            {\end{theorem}}

\newcommand\ba            {\begin{aligned}}
	\newcommand\ea            {\end{aligned}}

\usepackage[usenames,dvipsnames]{xcolor}

\usepackage{dcolumn}

\usepackage{pifont}
\usepackage{ulem}

\usepackage{multirow}

\def\({\left(}
\def\){\right)}
\def\[{\left[}
\def\]{\right]}
\newcommand{\ie}{\begin{equation}\begin{aligned}}
\newcommand{\fe}{\end{aligned}\end{equation}}
\def\d{\partial}

\begin{document}
\hfill MIT-CTP/5631

\title{Quantum vortex lattice: Lifshitz duality, topological defects and multipole symmetries}

\author{Yi-Hsien Du}
\email[Electronic address:$~~$]{yhdu@uchicago.edu}
\thanks{\\ These two authors contributed equally.}
\affiliation{Kadanoff Center for Theoretical Physics, \mbox{University of Chicago, Chicago, Illinois 60637 USA}}
\affiliation{Kavli Institute for Theoretical Physics, University of California, Santa Barbara, CA 93106, USA}

\author{Ho Tat Lam}
\email[Electronic address:$~~$]{htlam@mit.edu}
\thanks{\\ These two authors contributed equally.}
\affiliation{Center for Theoretical Physics, \mbox{Massachusetts Institute of Technology,  Cambridge, MA 02139 USA}}

\author{Leo Radzihovsky}
\email[Electronic address:$~~$]{radzihov@colorado.edu}
\affiliation{Department of Physics and Center for Theory of Quantum Matter, \mbox{University of Colorado, Boulder, CO 80309 USA}}

\date{\today}

\begin{abstract} 
We study an effective field theory of a vortex lattice in a two-dimensional neutral rotating superfluid. Utilizing particle-vortex dualities, we explore its formulation in terms of a $U(1)$ gauge theory coupled to elasticity, that at low energies reduces to a compact Lifshitz theory augmented with a Berry phase term encoding the vortex dynamics in the presence of a superflow. Utilizing elasticity- and Lifshitz-gauge theory dualities, we derive dual formulations of the vortex lattice in terms of a traceless symmetric scalar-charge theory and demonstrate low-energy equivalence of our dual gauge theory to its elasticity-gauge theory dual. We further discuss a multipole symmetry of the vortex lattice and its dual gauge theory's multipole one-form symmetries. We also study its topological crystalline defects, where the multipole one-form symmetry plays a prominent role. It classifies the defects, explains their restricted mobility, and characterizes descendant vortex phases, which includes a novel vortex supersolid phase. Using the dual gauge theory, we also develop a mean-field theory for the quantum melting transition from a vortex crystal to a vortex supersolid.
\end{abstract}

\maketitle

\tableofcontents

\section{Introduction}
Motivated by numerous direction -- robust quantum memory~\cite{PhysRevA.83.042330,RevModPhys.88.045005,PhysRevX.10.031041} and computing, beyond-quantum-field-theory classes of quantum liquids, quantum elasticity and melting~\cite{Beekman_2017,Pretko_2018,PhysRevLett.121.235301}, generalized symmetries~\cite{Gaiotto_2015,Qi_2021} -- recently there has been an explosion of research on models with quasi-particles characterized by restricted mobility~\cite{Nandkishore_2019,Pretko_2020,Du_2022,RevModPhys.96.011001,PhysRevResearch.6.L012040}. Significant progress was made in identifying a class of $U(1)$ fractonic gauge theories as duals of quantum crystals with topological defects~\cite{Pretko_2018,PhysRevLett.121.235301,PhysRevB.100.134113,RHvectorPRL2020}. This connection demystified and clarified some aspects of fractonic orders, provided their in-principle physical realization as crystalline topological defects, and uncovered generalized fractonic models as duals of well-understood elastic theories, such as e.g., quantum smectics~\cite{PhysRevLett.125.267601,Zhai_2021} and supersolids~\cite{PhysRevLett.121.235301,PhysRevB.100.134113,Kumar_2019}. Complementarily, such generalized gauge theories provided a compact effective field theory description of quantum crystals, their defects and associated quantum melting transitions. 

The duality was subsequently extended to time-reversal broken crystals such as for example a Wigner crystal~\cite{PhysRevB.100.134113,Kumar_2019} and rotating superfluids hosting a vortex lattice~\cite{Nguyen-Gromov-Moroz}. 
Although there is a strong similarity between resulting models and their fractonic tensor-gauge theory duals, a  vortex crystal is strongly coupled to its superfluid sector, that encodes superflow-mediated vortex lattice incompressibility, that on dualizing the bosonic sector, couples elasticity to a $U(1)$ vector gauge field. 

Upon a proliferation of mobile vortex vacancies and/or interstitials a vortex crystal is expected to undergo a quantum phase transition to a non-superfluid form, where the underlying superfluid order is destroyed ($U(1)$ symmetry of boson number is restored), a state that is the vortex lattice analog of a supersolid -- a superfluid crystal of bosons that exhibits ODLRO~\cite{PhysRevLett.121.235301,PhysRevB.100.134113}. Fundamentally, such ``normal'' (i.e., a non-superfluid) vortex lattice state is a rotated Mott insulating crystal of bosons. In this ``vortex supersolid'' the superflow is absent, with the aforementioned dual $U(1)$ gauge sector Higgs'ed and one expects the state to reduce to a time-reversal broken Wigner crystal with its guiding-center dynamics and previously studied gauge-theory dual~\cite{PhysRevB.100.134113,Kumar_2019}. We also expect that the crystal can further disorder through quantum melting into a smectic, hexatic or nematic vortex states~\cite{PhysRevLett.125.267601,Zhai_2021,RevModPhys.96.011001}.
These properties can be equivalently captured both in the direct and dual tensor-gauge theory descriptions~\cite{Nguyen-Gromov-Moroz}, though the analysis of phase transitions is particularly well accessible on the gauge-dual side, where they correspond to various types of generalized Higgs transitions~\cite{PhysRevLett.125.267601,Zhai_2021,RevModPhys.96.011001}. The classical analogs of such states have been previously studied in three-dimensional vortex states in type-II superconductors~\cite{FreyFisherNelsonSS,MarchettiLRSS}.

It was recently observed~\cite{PhysRevResearch.6.L012040} that in the superfluid vortex phase, the coupling to the superfluid flow ($U(1)$ vector gauge field in the dual description) leads to vortex lattice incompressibility. This constrains lattice distortions to be divergenceless, reducing its description to a generalized Lifshitz theory (studied previously in a number of other contexts~\cite{PhysRevB.65.024504,PhysRevB.69.224416,PhysRevB.69.224415,Ardonne_2004,Gorantla:2022eem,PhysRevB.106.064511,Gorantla:2022ssr}) of a single superfluid phase-like Goldstone mode. 
Duality of the Lifshitz theory has been analyzed in earlier works~\cite{Radzihovsky_2022,Gorantla:2022ssr} and we thus extend it here to duality of a vortex crystal.

\begin{figure}[h]
\centering
\includegraphics[width=0.48\textwidth]{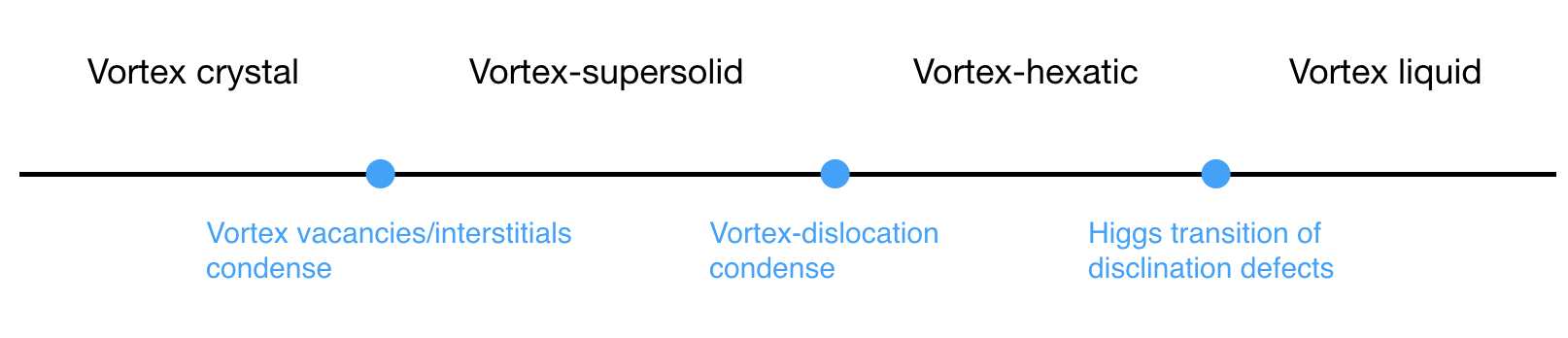} 
\caption{A schematic phase diagram for a vortex crystal and its associated phases studied in this paper. With increasing quantum fluctuations (e.g., rate of rotation), the vortex-crystal undergoes a quantum phase transition into a vortex-supersolid, a non-superfluid but periodic vortex state, driven by condensation of vortex
vacancies/interstitials. Upon further increase in quantum fluctuations, the vortex-supersolid undergoes a quantum melting transition to a vortex-hexatic (or, depending on microscopic interaction details, to a vortex-nematic or vortex-smectic), driven by condensation of vortex-dislocations~\cite{MarchettiLRSS, FreyFisherNelsonSS,PhysRevLett.125.267601,Zhai_2021}. At higher rotation, the system quantum melts into a vortex liquid phase.}\label{phasedia}
\end{figure}

We summarize the main results of this paper below.
First, we revisit the derivation of the low-energy effective Lifshitz theory for the vortex lattice, paying special attention to the compactness of the Lifshitz field. This allows us to uncover a rotation-induced Berry phase in the effective Lifshitz theory. Second, we derive a dual gauge theory description of the vortex lattice in terms of traceless symmetric scalar-charge theory. Third, using the effective Lifshitz theory and the dual gauge theory, we systematically analyze a hidden multipole (one-form) symmetry and the topological defects of the vortex lattice, which are associated with the winding of the Lifshitz field and the Wilson defects in the tensor gauge theory. Fourth, we characterize the neighboring quantum phases of the vortex lattice and formulate a mean-field theory for the transitions between the vortex lattice phase and these neighboring phases using the Higgs transitions of the dual gauge theory. 
These phases are illustrated schematically in Fig.~\ref{phasedia}.

This paper is organized as follows. 
In Sec.~\ref{sec:superfluid} we review superfluids and their $U(1)$ boson-vortex duality. 
We then introduce in Sec.~\ref{rotatedSFvortexlattice} a rotating superfluid and its linearized Lifshitz theory formulation. 
Then utilizing Lifshitz gauge duality, in Sec.~\ref{sec:duality} we describe vortex lattice in terms of traceless symmetric scalar-charge gauge theory and discuss its multipole symmetries in Sec.~\ref{sec:symmetry}, its defects and multipole one-form symmetries in Sec.~\ref{sec:defect}, and its descendant vortex phases and their phase transitions in Sec.~\ref{sec:phase}.

\section{Superfluid and its duality}
\label{sec:superfluid}
At low-energies, a superfluid is characterized by a local superfluid phase $\varphi({\bf x})\sim \varphi({\bf x})+2\pi$ and canonically conjugate boson density $n(\bf x)$, with boson operator $a\sim e^{i\varphi}$ satisfying $[e^{i\varphi},n] = e^{i\varphi}$. Its universal properties are well-captured by a Hamiltonian density (taking $\hbar = 1$):  
\begin{equation}
    {\cal{H}}_{sf} = \frac{n_s}{2} (\mathbf{\grad}\varphi)^2 + \frac{\chi^{-1}}{2} (\delta n)^2\;,
        \label{Hsf}
\end{equation}
with $\delta n=n-n_0$ the fluctuation of boson density $n$ away from its equilibrium value $n_0$, and $n_s$ and $\chi$ the superfluid stiffness and compressibility (proportional to the inverse of the boson interaction), respectively. 
A corresponding Lagrangian density is given by
\begin{equation}
    {\cal{L}}_{sf} =\frac{\chi}{2}(\partial_t\varphi)^2 -\frac{n_s}{2} (\mathbf{\grad}\varphi)^2\;.
    \label{Lsf}
\end{equation}
The bosonic particle 3-current is given by 
\begin{equation}
    J_\mu=(\delta n,J_i)_\mu = (\chi \partial_t\varphi,n_s\partial_i\varphi)_\mu\;,
    \label{bosonj}
\end{equation}
with the Euler-Lagrange equation encoding boson conservation via continuity relation: 
\begin{equation}
\partial^\mu J_\mu = 0\;,
\label{Jcontinuity}
\end{equation}
where $\mu=(t,x,y)$. 
In addition to these low-energy Goldstone mode excitations, a superfluid admits vortex excitations, corresponding to singular configurations of the compact phase $\varphi$: 
\begin{eqnarray}
\frac{1}{2\pi}\epsilon_{\mu\nu\gamma}\pt^\nu\pt^\gamma\varphi 
 = j_\mu\;,
\label{vorticity}
\end{eqnarray}
with the integer-valued vortex 3-current $j_\mu$, enabling superfluid rotation. Note that naively, the left-hand side of the equation vanishes due to the antisymmetric properties of the Levi-Civita tensor. However, it can be non-zero for singular, nonsingle-valued $\varphi$ configurations corresponding to vortices.

Following the standard boson-vortex
duality~\cite{peskin1978,DasguptaHalperin,PhysRevB.39.2756}, we express the bosonic Lagrangian \rf{Lsf} in terms of the bosonic current $J_\mu\equiv(\pt \times a)_\mu$, with $(\pt \times a)_\mu\equiv \epsilon_{\mu\nu\gamma}\pt^\nu a^\gamma$, whose continuity \rf{Jcontinuity} is solved by expressing it in terms of $U(1)$ gauge potential $a_\mu$. This thereby maps the bosonic 3-current $J$ onto a dual electromagnetic field, $b=\grad\times{\bf a}$, ${\bf e} = \pt_t {\bf a} - \grad a_t$: 
\begin{eqnarray}
J_\mu =(\delta n,J_i)_\mu\equiv \frac{1}{2\pi}(\delta b, \epsilon_{ij}e_j)_\mu,
\end{eqnarray}
with $\delta b = b - b_0$ (and $b_0/2\pi = n_0$, corresponding to the background boson density) and equivalently the Hodge-dual $^*J$ maps onto the electromagnetic field strength $f_{\mu\nu} = 2\pi \epsilon_{\mu\nu\gamma} J^\gamma$. With this, the bosonic continuity \rf{Jcontinuity} and circulation \rf{vorticity} equations map onto the Faraday and Ampere law, respectively, and the Lagrangian \rf{Lsf} to the Maxwell action:  
\begin{equation}
\cL = \frac{1}{8\pi^2 n_s} \mathbf{e}^2-\frac{1}{8\pi^2\chi}(\delta b)^2-a_\mu j^\mu\;,
\label{Ldual}
\end{equation}
coupled to charged matter representation of vortices.
Correspondingly, the Hamiltonian density \rf{Hsf}, transforms into $\cH = \frac{1}{8\pi^2 n_s} {\bf e}^2 + \frac{1}{8\pi^2\chi} (\grad\times \delta{\bf a})^2 - {\bf a}\cdot {\bf j}$, with  electric field $e_i$ and gauge vector potential $a_i$  canonically conjugate, and supplemented by a Gauss's law $\mathbf{\grad} \cdot\mathbf{e} = j_t$. The latter encodes vortex circulation density, corresponding to the temporal component of \rf{vorticity}.

\section{Rotating superfluid:~vortex lattice}
\label{rotatedSFvortexlattice}
An imposed rotation with angular velocity ${\bf\Omega}=\Omega\zh$ is conveniently treated in a rotating reference frame, where a particle of mass $m$ experiences the Coriolis force ${\bf F}=2m{\bf v}\times {\bf \Omega}$. It can be effectively viewed as the Lorentz force acting on a unit-charged particle moving in an effective magnetic field $B=2m\Omega$. At the level of a Lagrangian, $\Omega$ appears as a Lagrange multiplier $-{\bf \Omega}\cdot{\bf L} = -{\bf A}\cdot{\bf J}$, enforcing a nonzero angular momentum density ${\bf L} = n_sm\xv\times\grad\varphi$, with the effective external vector potential ${\bf A} = m{\bf \Omega}\times \xv$ inducing a particle current ${\bf J}$. 
In the dual language, the rotation is encoded via the external gauge potential ${\bf A}$ coupled to the particle current $\mathbf{J}=\frac{1}{2\pi}\mathbf{e}\times \mathbf{\hat z}$ leading to the coupling $-{\bf A}\cdot{\bf J}=\frac{1}{2\pi}A_i \epsilon_{ij} \partial_j a_t = \frac{1}{2\pi} a_t B$ (having integrated by parts under an implicit integral sign and dropped the $\partial_t a_i$ component of $e_i$ that is a total derivative for a time-independent $A_i$). As a consequence, a two-dimensional superfluid rotating at angular velocity $\Omega$ gives rise to the formation of a triangular lattice of quantized vortices with equilibrium vortex density:
\ie
n_v=\frac{2m\Omega}{2\pi}\equiv\frac{B}{2\pi}~.\label{nvB}
\fe
The elementary Goldstone-mode excitations of the vortex lattice, known as Tkachenko modes~\cite{tkachenko1965,tkachenko1966,tkachenko1969,Sonin_2014,PhysRevResearch.6.L012040,PhysRevB.109.035135}, possess several unconventional characteristics compared to ordinary sound waves in solids. Most notably, they are characterized by a quadratic dispersion $\omega \sim q^2$ at low momentum, with only one polarization.

\subsection{Effective vortex-lattice field theory}
\label{sec:EFT}
A recent discovery has revealed that the Tkachenko modes can be described by a non-linear Lifshitz scalar theory, which incorporates non-linear dipole and higher multipole symmetries arising from magnetic translations and rotations~\cite{PhysRevResearch.6.L012040} and will be the starting point of our analysis. The non-linear theory of the Tkachenko mode allows a convenient formulation as a noncommutative field theory~\cite{Douglas:2001ba,Rubakov-NC,du2021noncommutative}. 
Here, we begin with a review of this effective field theory of a vortex lattice and then derive the linearized Lagrangian by expanding the effective field theory to quadratic order~\cite{PhysRevResearch.6.L012040}.

As any two-dimensional crystal, a vortex lattice can be described by two scalar fields frozen into the lattice $X^a(t, x^i)$ with $a=x,y$, which is related to the lattice displacement $u^a$ by $X^a(t,x^i)=\delta_i^a x^i-u^a(t,x^i)$, in the Eulerian (laboratory) coordinates $x^i$. The Lagrangian density consists of the elastic component:
\be
\mathcal{L}_{el}=\frac{1}{2}\rho(\partial_t u_i)^2 - \oh K_{ij;kl} u_{ij}u_{kl}\label{Lel}
\ee
and the superfluid sector $\cL_{sf}$, \rf{Lsf}, coupled  through the superflow ($J_\mu$) and vortex ($j_\mu$) currents. In \rf{Lel},  $\rho$ is the effective vortex mass density, $K_{ij;kl}$ is the elastic tensor, and $u_{ij}=\frac{1}{2}(\partial_i u_j+\partial_j u_i)$
is the linearized symmetric strain tensor.

The crucial coupling of the two sectors is most transparently implemented in the superfluid's gauge-dual description \rf{Ldual}, via $-a_\mu j^\mu$, with the vortex current given by 
\ie
j_\mu=\frac{1}{2}n_v\epsilon_{\mu\nu\gamma}\epsilon_{ab}\partial^\nu X^a\partial^\gamma X^b ~ ,
\fe
that is a ``space-time Jacobian'' between the reference and vortex lattice coordinates. In terms of the phonon distortions, the temporal and spatial currents take the form:
\ie\label{eq:vortex_current}
&j_t=n_v-n_v\d_i u^i+\frac{1}{2} n_v\epsilon_{ij}\epsilon_{ab}\d_i u_a\d_j u_b~,
\\
&j_i=-n_v\d_tu_i-n_v\epsilon_{ij}\epsilon_{ab}\d_j u_a\d_t u_b~,
\fe
that have a natural and transparent interpretation. 
Combining these inside $-a_\mu j^\mu$, together with the elastic $\cL_{el}$ and electromagnetic (superfluid) $\cL_{sf}$ sectors and the rotation-imposed external vector potential, $-{\bf A}\cdot{\bf J}$, we obtain the vortex lattice Lagrangian density, that, at harmonic level and dropping the subdominant kinetic $\oh\rho (\partial_t u_i)^2$ and electric $\frac{1}{8\pi^2n_s}{\bf e}^2$ energies takes the form:
\ie\label{eq:micro_Lagrangian}
    \mathcal{L}=&- \mu u_{ij}^2 - \frac{\lambda}{2}u_{ii}^2- \frac{1}{8\pi^2\chi}(\delta b)^2 
    \\
    &+a_t\left(\frac{B}{2\pi} - n_v\right) + n_v e_i u_i
   +\frac{1}{2}n_v b_0\epsilon_{ij}u_i\partial_t{u}_j\;. 
\fe
Above, for simplicity we have taken isotropic elasticity with elastic tensor:
\ie\label{eq:elastic-tensor}
K_{ij;kl}=\mu (\delta_{ik}\delta_{jl}+\delta_{il}\delta_{jk})+\lambda \delta_{ij}\delta_{kl}~,
\fe
characterized by two Lam\'e elastic constants, $\mu$ (shear modulus) and $\lambda$ (valid for a triangular lattice) and integrated by parts in the second and last terms of the second line. We also restricted our analysis to a harmonic approximation, with $b$ replaced by $b_0=2\pi n_0$ in the last term, set by the equilibrium particle number density $n_0$.
This last term crucially encodes guiding center vortex dynamics, that sees background boson density $n_0$ as an effective magnetic field $b_0 = 2\pi n_0$, equivalently corresponding to the Magnus force on a moving vortex. It encodes noncommutativity of components of the phonon operator, but with the commutation relation modified by the superflow encoded in the gauge field $a_i$. 
The Lagrangian explicitly breaks the spatial reflection ($\mathbf{R}$) and time-reversal ($\mathbf{T}$) symmetry  but is still invariant under the combined $\mathbf{RT}$ transformation~\cite{Moroz_2018,Moroz_2019,PhysRevResearch.6.L012040}.

Functionally integrating over $a_t$ in the partition function, locks the equilibrium vortex density $n_v$ to the flux density $B$ (equivalently to the imposed angular velocity $\Omega$, with critical velocity $\Omega_{c}$ vanishing in the thermodynamic limit) with one unit of flux quantum per vortex lattice unit cell according to \rf{nvB}.   At long wavelengths, this then introduces a crucial incompressibility constraint on the vortex lattice:
\ie\label{eq:EOM1}
\d_i u_i=0~,
\fe
forcing a transversal form of the vortex lattice phonon $u_i(\xv)$. We emphasize that this incompressibility constraint is for the vortex lattice, not the underlying superfluid~\footnote{For an incompressible fluid, the dispersion is linear instead of quadratic. It can be understood as follows. For vanishing fluid compressibility, $\chi=0$ and the kinetic term $\frac{\chi}{2}(\partial_t\phi)^2$ in the effective Lifshitz theory \eqref{eq:phi_Lagrangian} vanishes. It is then necessary to retain the subdominant term $\frac{\rho}{2}(\partial_t u_i)^2$ in \eqref{Lel}, which under \eqref{eq:u_to_phi} becomes $\frac{\rho}{2B^2}(\partial_t \partial_i \phi)^2$. The dispersion is then modified to a linear one.}. We also note that keeping the subdominant electric energy term $\frac{1}{8\pi^2n_s}{\bf e}^2$ in the above Lagrangian, would allow for nonzero vortex lattice comprehensibility. At long scales, the incompressibility constraint reduces the Lagrangian \rf{eq:micro_Lagrangian} to 
\ie\label{eq:micro_Lagrangian2}
    \mathcal{L}=&- \mu u_{ij}^2 - \frac{1}{8\pi^2\chi}(\delta b)^2 
    - n_v a_i \pt_t u_i
   +\frac{1}{2}n_v b_0\epsilon_{ij}u_i\partial_t{u}_j\;,
\fe
with the Lam\'e elastic constant $\lambda$ dropping out due to the incompressibility condition \eqref{eq:EOM1}.  It is straightforward to see (via, e.g., the equation of motion) that \rf{eq:micro_Lagrangian2} encodes a single-polarization Tkachenko modes with a quadratic dispersion relation $\omega \sim q^2$. We emphasize, that, because of the Berry phase coupling of phonon velocity $\pt_t u_i$ to the superfluid gauge-field $a_i$, the commutation relations are distinct from that of a pure guiding-center dynamics (in the absence of the gauge field) of e.g., a Wigner crystal in a magnetic field, though the quadratic dispersion is the same.  

The Euler-Lagrange equation for $u_i$ is given by
\ie\label{eq:EOM2}
n_ve_i+n_v b_0\epsilon_{ij}\partial_t{u}_j+2 \mu\partial_j u_{ij} +\lambda\partial_i u_{kk}=0~.
\fe
It encodes a local equilibrium force balance in the vortex lattice. The first and second terms are the dual ``electric'' and ``magnetic'' forces (dual ``Lorentz'' force), corresponding to a Magnus force experienced by a moving vortex due to a superflow $J_i = \frac{1}{2\pi}\epsilon_{ij} e_j$. It is equivalent to Kelvin's condition that in the absence of other forces, a vortex moves with a local superfluid velocity. The third and fourth terms represent a force associated with elastic stresses. 
Equations \eqref{eq:EOM1} and \eqref{eq:EOM2} will be important for our subsequent analyses.

We take the vortex lattice to be triangular, corresponding to the expected lowest energy state for repulsive vortex interaction. A triangular unit cell has area $\sqrt{3}l^2/2$, with the equilibrium vortex density $n_v$ related to the lattice constant $l$ by $n_v={2}/{\sqrt{3}l^2}$. Our discussions so far have neglected vortex lattice topological defects -- disclinations, dislocations, vacancies and interstitials. These enter the low-energy linearized effective field theory \eqref{eq:micro_Lagrangian} through the compactness of $u_i$, associated with lattice periodicity and captured by the identifications on the displacement field $u_i$:
\ie\label{eq:identi_translation}
u_x\sim u_x+k_xl+\frac{1}{2}k_yl~,
\quad u_y\sim u_y+\frac{\sqrt{3}}{2}k_yl~,
\fe
where $k_i$ are integers. Similarly, a $2\pi/6$ rotation around a lattice site leaves the lattice invariant so the bond angle $\theta$ should be $2\pi/6$-periodic. In the linearized elasticity theory, the bond angle is approximated by $\theta\approx\frac{1}{2}\epsilon_{ij}\partial_i u_j$ which together with the $2\pi/6$-periodicity of the bond angle implies the following identification on $u_i$:
\ie\label{eq:identi_rotation}
u_i\sim u_i+\frac{2\pi k}{6}\epsilon^{ij}x_j~,
\fe
where $k$ is an integer. The above thereby allow singular configurations, with nonzero $\grad\times\grad u_i$ and $\grad\times\grad\theta$, which will be vitally important for the defect analysis in Sec.~\ref{sec:defect}. 

\subsection{Compact Lifshitz theory with a Berry phase}

The Lifshitz field theory arises from the linearized effective field theory \eqref{eq:micro_Lagrangian2} of the vortex lattice by integrating out the $U(1)$ vector gauge field $a_\mu$ representing the bosonic matter component and the associated superflow. 
As we saw in the previous section, the integration over $a_t$ imposed the incompressibility condition \eqref{eq:EOM1} and transformed \eqref{eq:micro_Lagrangian} into \eqref{eq:micro_Lagrangian2}. The condition \eqref{eq:EOM1} is straightforwardly solved using a phase $\phi(\xv)$, defined by
\ie\label{eq:u_to_phi}
u_i=\frac{1}{B}\epsilon_{ij}\partial_j \phi~.
\fe
A priori, at this stage, $\phi(\xv)$ is unrelated to the superfluid phase $\varphi(\xv)$ of the previous section. However, as we will see below, $\phi(\xv)$ is a deviation from $\varphi_{VL}(\xv)$ describing a triangular vortex lattice of single vortex configurations, $\varphi_{vortex}(\xv)=\arctan[(y-y_i)/(x-x_i)]$. 
The linearized symmetric strain tensor can then be expressed as
\ie\label{eq:strain_tensor_in_u}
u_{ij}=\frac{1}{2}\left(\partial_i u_j+\partial_j u_i\right)=\frac{1}{B}\epsilon_{jk}D_{ik}\phi~,
\fe
where we defined a differential operator:
\ie
D_{ij}\phi\equiv \left( \partial_i \partial_j -\frac{1}{2}\delta_{ij}\partial^2 \right) \phi~.
\fe
Next, integrating out $a_i$ gives
\ie\label{eq:b=d_tphi}
\frac{1}{2\pi} \delta b=-\chi\partial_t\phi~,
\fe
which together with \eqref{eq:u_to_phi} reduces the Lagrangian \eqref{eq:micro_Lagrangian2} to a {\it linearized} (quadratic) Lifshitz theory:
\ie\label{eq:phi_Lagrangian}
\mathcal{L}= \frac{\chi}{2}\left(\partial_t \phi\right)^2-\frac{\mu}{ B^2}\left(D_{ij}\phi\right)^2
+\frac{b_0}{4\pi B}\epsilon_{ij} \d_i\phi\d_j \d_t\phi~,
\fe
supplemented by a Berry phase, that is a distinctive characteristic of a quantum vortex lattice, crucially contrasting it from the prevously-studied Lifshitz theory description \cite{Watanabe:2013iia}.
The Berry phase is the only term in the Lagrangian that breaks the $\mathbf{T}$ and $\mathbf{R}$ symmetry while preserving the combined $\mathbf{RT}$ symmetry. 

Naively, one may hastily neglect the Berry phase term, arguing that as a total derivative, it is unimportant in the bulk.  
However, since $\phi(\xv)$ is a compact angular field, it can wind, encoding singular vortex configurations for which  $\frac{1}{2\pi}\epsilon_{ij}\d_i\d_j\phi = \frac{1}{2\pi}\grad\times\grad\phi =\delta j_t(\xv)=j_t(\xv) - j_t^{VL}(\xv)\neq 0$, where $j_t^{VL}(\xv)=\grad\times\grad\varphi_{VL}$ is the vortex density corresponding to a vortex lattice in $\varphi(\xv)$, with average $n_v$.  For such vortex configurations the Berry phase term contributes nontrivially in the bulk:
\bea\label{eq:LBerry}
\mathcal{L}_{{Berry}}
=\frac{b_0}{4\pi B}\epsilon_{ij} (\d_i\d_j\phi)\d_t\phi=\frac{1}{2}\nu\delta j_t\d_t\phi~,
\eea
where $\nu\equiv b_0/B$ is the filling fraction of boson per vortex. 

Not unrelated, it is tempting to rewrite the second ``elastic" term as simply $\frac{\mu}{2B^2}(\nabla^2 \phi)^2$ through integration by parts. However, in the presence of vortices there is an obstruction to integration by parts as derivatives on $\phi_{vortex}$ do not commute (see \cite{Pretko_2018,PhysRevLett.121.235301,PhysRevB.100.134113,Gorantla:2022ssr} for related discussions). 

The compactness of $\phi$ follows from the identifications on $u_i$, listed in \eqref{eq:identi_translation} and \eqref{eq:identi_rotation}, and the relationship between $u_i$ and $\phi$ in \eqref{eq:u_to_phi}, which lead to the following identifications:
\ie\label{eq:phi_identification}
\phi(\xv)\sim\phi(\xv)+k_xyBl +\frac{1}{2}k_y(y-\sqrt{3}x)Bl+\frac{2\pi  k}{12}Br^2~,
\fe
with $r=\sqrt{x^2+y^2}$ the radial coordinate and integer $k_x,k_y,k$.
In addition to these, there is another constant identification on $\phi$:
\ie\label{eq:2pi-phi}
\phi(\xv)\sim \phi(\xv)+2\pi~,
\fe
following from the $2\pi$ periodicity of the superfluid phase $\varphi(\xv)\sim\varphi(\xv)+2\pi$.

The Tkachenko modes are made explicit in the linearized Lifshitz theory \rf{eq:phi_Lagrangian}. They are the plane wave excitations of the phase field $\phi$, that manifestly have only one polarization and a quadratic dispersion with dynamical exponent $z=2$, characteristic of the Lifshitz symmetry $(t,x_i) \rightarrow (\xi t,\xi^2 x_i)$, with the scaling factor $\xi$~\cite{Note1}. 

For later use, we summarize the operator maps between the effective field theory \eqref{eq:micro_Lagrangian2} and  the linearized Lifshitz theory \eqref{eq:phi_Lagrangian}:
\begin{equation}\label{eq:op_map}
 \renewcommand{\arraystretch}{2.1}
 \begin{array}{ccc}
 u_i\quad&\quad\longleftrightarrow\quad &\quad\dfrac{1}{B}\epsilon_{ij}\d_j \phi~,
 \\
 u_{ij}\quad&\quad\longleftrightarrow\quad &\quad\dfrac{1}{B}\epsilon^{jk}D_{ik} \phi~,
 \\
 \dfrac{1}{2\pi}\delta b\quad &\quad\longleftrightarrow\quad &\quad-\chi\d_t\phi~,
 \\
 \dfrac{1}{2\pi}e_i\quad &\quad\longleftrightarrow\quad  &\quad\dfrac{b_0}{2\pi B}\d_i\partial_t \phi- \dfrac{\mu}{B^2}\epsilon_{ij}\partial_j\partial^2 \phi~.
 \end{array}
 \end{equation}
The last relation can be derived by substituting \eqref{eq:u_to_phi} into the Euler-Lagrange equation \eqref{eq:EOM2}.

\section{Duality}
\label{sec:duality}
In this section we derive a number of equivalent formulations of the vortex lattice starting from the Lifshitz field theory presented in \eqref{eq:phi_Lagrangian} and the effective field theory presented in \eqref{eq:micro_Lagrangian2}. 
We apply a generalized Lifshitz duality \cite{Gorantla:2022ssr,Radzihovsky_2022} to the Lifshitz field theory~\cite{PhysRevResearch.6.L012040}, obtaining the main new result of this section -- a \textit{traceless symmetric scalar-charge gauge theory} description of the vortex lattice. We then demonstrate the low-energy equivalence of this dual gauge theory to the vortex lattice description via fracton-elasticity duality \cite{Pretko_2018,PhysRevLett.121.235301,PhysRevB.100.134113}, derived in an earlier work \cite{Nguyen-Gromov-Moroz}. 

\subsection{Lifshitz duality to traceless symmetric scalar-charge gauge theory}\label{sec:Lifshtiz-duality}
We now derive another equivalent description of the vortex lattice by dualizing the linearized Lifshitz theory to a traceless symmetric scalar-charge gauge theory, generalizing the Lifshitz duality studied in \cite{Gorantla:2022ssr,Radzihovsky_2022}. 

To this end we introduce Hubbard-Stratonovich (HS) fields $\hat\pi_t,\hat \pi_{ij}$ and $u_i$ into Lagrangian $\cL[\phi]$ \eqref{eq:phi_Lagrangian} so that $\phi$ appears linearly in the transformed Lagrangian: 
\ie\label{eq:Lagrangian-tensor-gauge-theory-1}
\mathcal{L}[\phi,\hat\pi,u_i]=&-\frac{1}{2\chi}\hat \pi_t^2+\hat \pi_t\partial_t\phi
+\frac{B^2}{4\mu}\hat \pi_{ij}^2+\hat\pi_{ij} D_{ij}\phi
\\
&-\frac{b_0 B}{4\pi }\epsilon_{ij}  u_i \partial_t u_j-\frac{b_0}{2\pi }u_i\partial_t\partial_i\phi~,
\fe
where $\hat \pi_{ij}$ is a {traceless} symmetric tensor, with a hat symbol emphasizing its tracelessness. 
The original Lifshitz Lagrangian is straightforwardly recovered, integrating out in the path integral the $\hat\pi_t$, $\hat\pi_{ij}$, and $u_i$ fields. At the harmonic level, this establishes relations:
\ie\label{eq:pi-phi}
\hat \pi_t=\chi\partial_t\phi~,\quad \hat\pi_{ij}=-\frac{2\mu}{B^2}D_{ij}\phi~,\quad u_i=\frac{1}{B}\epsilon_{ij}\partial_j\phi~,
\fe
which reveals that $\hat\pi_t$, $\hat \pi_{ij}$, and $u_i$ play the roles of the boson density, vortex lattice stress, and the displacement (phonon) field, respectively.
We note that, although we used the same symbol $u_i$, there was no a priori relationship with the vortex lattice phonon field appearing in \rf{Lel}. However, the third equation above, relating $u_i$ and $\phi$ is exactly the one that appeared in \eqref{eq:u_to_phi} thereby identifying HS field $u_i$ with the vortex displacement field.

To obtain the dual Lagrangian, we integrate out $\phi$, that appears linearly and thereby gives the following constraint:
\ie
\partial_t\hat\pi_t-D_{ij}\hat \pi_{ij}+\frac{b_0}{2\pi}\partial_t\partial_i u_i=0~.
\fe
We note that through the last term, the dynamics here is qualitatively modified by the Berry phase, encoding a coupling between superfluid current $\pt_i\phi$, vortex density $~\pt_i u_i$ and boson density $\hat\pi_t$, that physically goes back to Kelvin's condition for vortex motion in a superflow.  To solve the constraint, it is convenient to define dual magnetic and electric fields (that respectively encode a sum of boson and vortex densities, and vortex-lattice stress):
\ie\label{eq:fieldstrength-pi}
\hat{\mathcal{B}}\equiv 2\pi \hat\pi_t+b_0\partial_iu_i~,
\quad\hat{\mathcal{E}}_{ij}\equiv 2\pi \epsilon_{jk}\hat\pi_{ik}~,
\fe
in terms of which constraint transforms into Faraday-like law:
\ie
\partial_t\hat{\mathcal{B}}-\epsilon_{ik}D_{jk}\hat{\mathcal{E}}_{ij}=0~.
\fe
Here, $\hat{\mathcal{E}}_{ij}$ is a traceless symmetric tensor following from the fact that $\hat \pi_{ij}$ is traceless symmetric, respectively. The constraint can be solved by introducing a tensor gauge field:
\ie\label{eq:traceless-gauge-field-maintext}
\hat{\mathcal{A}}_t\sim \hat{\mathcal{A}}_t+\partial_t \hat\lambda~,
\quad \hat{\mathcal{A}}_{ij}\sim \hat{\mathcal{A}}_{ij}+D_{ij} \hat\lambda~,
\fe
with $\hat{\mathcal{A}}_{ij}$ a traceless symmetric tensor
and identifying $\hat{\mathcal{E}}_{ij}$, $\hat{\mathcal{B}}$ with the gauge-invariant field strengths of the tensor gauge field:
\ie
\hat{\mathcal{E}}_{ij}=\d_t  \hat{\mathcal{A}}_{ij} -  D_{ij}\hat{\mathcal{A}}_t~, \quad \hat{\mathcal{B}}=
\epsilon_{ik}D_{jk} \hat{\mathcal{A}}_{ij}~.
\fe
This then gives our main result, the dual traceless tensor gauge theory Lagrangian:
\ie\label{eq:Lagrangian-tensor-gauge-theory-2}
\mathcal{L}=\frac{B^2}{16\pi^2\mu}\hat{\mathcal{E}}_{ij}^2-\frac{1}{8\pi^2\chi}(\hat{\mathcal{B}}-b_0\partial_iu_i)^2
-\frac{b_0 B}{4\pi }\epsilon_{ij}  u_i \partial_t u_j~.
\fe
As $u_i$ appears quadratically in the Lagrangian, we can further integrate it out and reduce the Lagrangian to a functional depending only on the traceless symmetric tensor gauge fields. However, because $u_i$ is gapless, this will result in a long-range interaction and thus will obscure the locality of the resulting Lagrangian. Thus, we keep the $u_i$ field and present the Lagrangian in the current form. 
In Appendix~\ref{app:duality-no-berry}, for comparison, we study the duality of the Lifshitz theory \eqref{eq:phi_Lagrangian} without the Berry phase.

This tensor gauge theory description of the vortex lattice has the advantage of making the topological crystalline defects and their mobility explicit (see Sec.\ \ref{sec:defect}). It will also serve as the starting point for our study of the vortex phases that can appear after melting of the vortex lattice.

For later use, we summarize the operator maps between the Lifshitz theory \eqref{eq:phi_Lagrangian} and the tensor gauge theory \eqref{eq:Lagrangian-tensor-gauge-theory-2}:
\ie\label{eq:op_map_tensor_gauge_theory}
 \renewcommand{\arraystretch}{2.1}
 \begin{array}{ccc}
 \partial_t\phi &\quad \longleftrightarrow\quad &\quad \dfrac{1}{2\pi\chi}(\hat{\mathcal{B}}-b_0\partial_i u_i)~,\\
D_{ij}\phi &\quad \longleftrightarrow\quad &\quad \dfrac{B^2}{4\pi\mu}\epsilon_{jk}\hat{\mathcal{E}}_{ik}~,
 \end{array}
 \fe
that follow from \eqref{eq:pi-phi} and \eqref{eq:fieldstrength-pi}.

\subsection{Vortex-lattice dual through fracton-elasticity duality}

In a distinct approach, authors of Ref.~\onlinecite{Nguyen-Gromov-Moroz} dualized a vortex lattice by applying fracton-elasticity duality~\cite{Pretko_2018,PhysRevLett.121.235301,PhysRevB.100.134113} to the elastic sector of the effective field theory \eqref{eq:micro_Lagrangian}, obtaining a symmetric scalar charge theory coupled to a $U(1)$ vector gauge field of the superfluid. We review their derivation below and show that our traceless symmetric tensor gauge theory \eqref{eq:Lagrangian-tensor-gauge-theory-2} consistently emerges in the low energy limit.

Following fracton-elasticity duality~\cite{Pretko_2018,PhysRevLett.121.235301,PhysRevB.100.134113}, we linearize the appearance of the phonon field $u_i$ in  \eqref{eq:micro_Lagrangian} by respectively introducing the lattice momentum, stress tensor and bond angle HS fields $\pi_i,\sigma_{ij},\theta$, that transforms the effective field theory to
\ie\label{eq:Lagrangian-FEdual-1}
\mathcal{L}=&\,\frac{1}{2}K_{ij;kl}^{-1}\sigma_{ij}\sigma_{kl}+\sigma_{ij} (\partial_i u_j-\theta\epsilon_{ij})
\\
&\,-\frac{1}{2n_vb_0}\epsilon_{ij}\pi_i\partial_t\pi_j+\pi_i\partial_t u_i+n_v e_i u_i-\frac{1}{8\pi^2\chi}(\delta b)^2
~ ,
\fe
with $K_{ij;kl}^{-1}$ the inverse of the elastic tensor \eqref{eq:elastic-tensor}:
\ie
K_{ij;kl}^{-1}=\frac{2\mu \delta_{ik}\delta_{jl}+\lambda(2\delta_{ik}\delta_{jl}-\delta_{ij}\delta_{kl})}{4\mu(\lambda+\mu)}~.
\fe
The role of the bond angle field $\theta$ is to relax the symmetrization of the strain tensor $u_{ij}$ to $\partial_i u_j-\theta\epsilon_{ij}$, with $\theta$ effectively Higgs'ing the antisymmetric component of the unsymmetrized strain $\partial_i u_j$~\cite{RHvectorPRL2020}.
The original Lagrangian is recovered by integrating over $\theta,\pi_i, \sigma_{ij}$ in the partition function, which, because the model is quadratic, gives the relations:
\ie\label{eq:pi-to-u}
\pi_i=-n_v b_0\epsilon_{ij} u_j~,\quad \sigma_{ij}=-K_{ij;kl}u_{kl}~,\quad \theta=\frac{1}{2}\epsilon_{ij}\partial_i u_j~.
\fe
To derive the dual Lagrangian, we instead integrate out (exactly) $u_i$ and $\theta$, which gives rise to the following constraints:
\ie
\partial_t \pi_i+\partial_j \sigma_{ji}=n_v e_i~,\quad \epsilon_{ij}\sigma_{ij}=0~.
\fe
The first constraint is the continuity equation for the vortex lattice stress-energy tensor broken by a dual electric field associated with the superflow, that exerts a force on the lattice. The second constraint symmetrizes the stress tensor $\sigma_{ij}$.
To solve the constraint, it is convenient to define dual magnetic and electric fields:
\ie\label{eq:define-Bi-Eij}
\mathcal{B}_i&\equiv-\frac{1}{n_v}\epsilon_{ij}(\pi_j-n_v \delta a_j)~,
\\
\mathcal{E}_{ij}&\equiv-\frac{1}{n_v}\epsilon_{ik}\epsilon_{jl}(\sigma_{kl}+n_v \delta a_t\delta_{kl} )~,
\fe
with $\delta a_\mu = a_\mu-{a}_{0,\mu}$ the fluctuation around the equilibrium gauge field $a_{0,\mu}$, whose magnetic field is $b_0$ and electric field vanishes.
In terms of these new fields, the constraint reduces to a Faraday-like law~\cite{Pretko_2018,PhysRevLett.121.235301,PhysRevB.100.134113,RHvectorPRL2020} and symmetrization of the electric field tensor:
\ie
\partial_t\mathcal{B}_i-\epsilon_{jk}\partial_j \mathcal{E}_{ki}
=0~,\quad \epsilon_{ij}\mathcal{E}_{ij}=0~.
\fe
These can be solved by introducing a symmetric tensor gauge field:
\ie\label{eq:define-sym-gauge-field}
\mathcal{A}_t\sim \mathcal{A}_t+\partial_t\lambda~,\quad
\mathcal{A}_{ij}\sim \mathcal{A}_{ij}+\partial_i\partial_j \lambda~,
\fe
and setting $\mathcal{B}_i$, $\mathcal{E}_{ij}$ to be its gauge-invariant field strengths:
\ie
\mathcal{B}_i=\epsilon_{jk} \partial_j \mathcal{A}_{ki}~,
\quad
\mathcal{E}_{ij}=\partial_t \mathcal{A}_{ij}-\partial_i\partial_j \mathcal{A}_t~.
\fe
We emphasize that, unlike \eqref{eq:traceless-gauge-field-maintext}, this symmetric tensor gauge field is not traceless.
With this, utilizing this gauge field description, we arrive at the dual Lagrangian derived in Ref.~\cite{Nguyen-Gromov-Moroz}: 
\ie\label{eq:Lagrangian-FE-final}
\mathcal{L}=&\,\frac{1}{2}n_v^2\tilde K^{-1}_{ij;kl}(\mathcal{E}_{ij}+ \delta a_t\delta_{ij})(\mathcal{E}_{kl}+ \delta a_t\delta_{kl})
\\
&\,-\frac{B}{4\pi b_0}\epsilon_{ij}(\mathcal{B}_i-\epsilon_{ik}\delta a_k)\partial_t(\mathcal{B}_j
\\
&\,-\epsilon_{jl}\delta a_l)-\frac{1}{8\pi^2\chi}(\delta b)^2~,
\fe
with $\tilde K^{-1}_{ij;kl}=\epsilon_{ii'}\epsilon_{jj'}\epsilon_{kk'}\epsilon_{ll'}K^{-1}_{i'j';k'l'}$.
It describes a symmetric tensor gauge theory coupled to a vector gauge theory.
For the Lagrangian to be gauge invariant, $\mathcal{A}_{ij}$ must transform under the gauge redundancy of $\delta a_\mu$ as
\ie\label{eq:beta-gauge-symmetry}
\delta a_\mu\sim \delta a_\mu+\partial_\mu \beta~,\quad
\mathcal{A}_{ij}\sim \mathcal{A}_{ij}+\beta\delta_{ij}~,
\fe
in additional to its own gauge redundancy \eqref{eq:define-sym-gauge-field}.

We now relate this description of the vortex lattice with our {\it traceless} symmetric tensor gauge theory \eqref{eq:Lagrangian-tensor-gauge-theory-2}. To this end we define the following fields identifications:
\ie\label{eq:define}
u_i&\equiv \frac{1}{b_0}( \mathcal{B}_i-\epsilon_{ij}\delta a_j)~,
\\
(\hat{\mathcal{A}}_t,
\hat{\mathcal{A}}_{ij})&\equiv \Big(\mathcal{A}_t,\mathcal{A}_{ij}-\frac{1}{2}\delta_{ij} \mathcal{A}_{kk}\Big)~.
\fe
Here, $u_i$ is the displacement field with the definition consistent with \eqref{eq:pi-to-u} and \eqref{eq:define-Bi-Eij}, and $(\hat{\mathcal{A}}_t, \hat{\mathcal{A}}_{ij})$ is a traceless symmetric tensor gauge field which share the same gauge redundancy as \eqref{eq:traceless-gauge-field-maintext} with the gauge parameter $\hat\lambda=\lambda$ and is invariant under the gauge transformation \eqref{eq:beta-gauge-symmetry} associated with $\beta$. The field strength of $(\hat{\mathcal{A}}_t,\hat{\mathcal{A}}_{ij})$ is related to the field strength of $(\mathcal{A}_t,\mathcal{A}_{ij})$ by
\ie
\hat{\mathcal{E}}_{ij}=\mathcal{E}_{ij}-\frac{1}{2}\delta_{ij}\mathcal{E}_{kk}~,\quad
\hat{\mathcal{B}}=-\partial_i\mathcal{B}_i~.
\fe
With this, we can reduce the Lagrangian \eqref{eq:Lagrangian-FE-final} into a functional of these new fields.
To this end, we integrate out $\delta a_t$ which sets a relation:
\ie\label{eq:a_t_EOM}
\delta a_t=-\frac{1}{2}\mathcal{E}_{kk}~.
\fe
The first term of the Lagrangian \rf{eq:Lagrangian-FE-final} then transforms into
\ie\label{eq:E-term}
\frac{1}{2}n_v^2\tilde K^{-1}_{ij;kl}\hat{\mathcal{E}}_{ij}\hat{\mathcal{E}_{kl}}=\frac{B^2}{16\pi^2\mu}\hat{\mathcal{E}}_{ij}\hat{\mathcal{E}}_{ij}~,
\fe
which simplifies significantly due to the tracelessness of $\hat{\mathcal{E}}_{ij}$.
Expressing the remaining components of the Lagrangian in terms of the newly defined fields gives
\ie
-\frac{b_0B}{4\pi }\epsilon_{ij}u_i\partial_tu_j-\frac{1}{8\pi^2\chi}(\hat{\mathcal{B}}-b_0\partial_i u_i)^2~,
\fe
which together with \eqref{eq:E-term},  reduces to our Lagrangian \eqref{eq:Lagrangian-tensor-gauge-theory-2} of the {\it traceless} symmetric tensor gauge theory.

\section{Global symmetries}\label{sec:symmetry}

\subsection{Microscopic symmetries}

In this section, we analyze the global symmetries of the effective field theory \eqref{eq:micro_Lagrangian} and track them across dualities. These global symmetries include the $U(1)$ symmetry associated with boson number conservation and the translation and rotation symmetry of the vortex lattice.
 
The microscopic boson number conservation is realized in the effective field theory \eqref{eq:micro_Lagrangian} as a $U(1)$ symmetry generated by the current:
\ie
J_\mu=\frac{1}{2\pi}\epsilon_{\mu\nu\rho}\partial_\nu a_\rho-n_0 \delta_{\mu,t}
~.
\fe
In our definition above, we subtracted the equilibrium boson density $n_0$ in the temporal component of the current $J_t$, so that $J_t$ represents the {\em fluctuation} of the boson density around its equilibrium value. This current obeys a continuity equation:
\ie
\partial^\mu J_\mu=0~,
\fe
corresponding to the local boson number conservation.
The quantized conserved charge is 
\ie\label{eq:particle_charge}
Q=\int d^2x\ J_t~,
\fe
which measures the fluctuations of the boson number around its equilibrium value.

The translation and rotation symmetry of the vortex lattice acts on the displacement fields $u_i$ as 
\ie
u_i\rightarrow u_i+c_x\delta_{i,x}+\frac{1}{2}c_y(\delta_{i,x}+\sqrt{3}\delta_{i,y})+\alpha\epsilon^{ij}x_j~,
\fe
where $c_x$ and $c_y$ are respectively the translations along the $x$ axis and along the axis $2\pi/6$ counter-clockwise to the $x$ axis, and $\alpha$ is the rotation angle.
We parameterized the symmetry transformation in such a way so that the  parameters have the periodicity $c_i\sim c_i+l$ and $\alpha\sim \alpha+2\pi/6$. The rotation symmetry parametrized by $\alpha$ takes the form of an infinitesimal rotation, although it remains a symmetry of the effective field theory \eqref{eq:micro_Lagrangian} at nonzero $\alpha$ due to linearized-elasticity of the Lagrangian. We note that these symmetries should not be confused with the translation and rotation symmetry of the full system.

The current $J_{\mu\nu}$ associated with translational and rotational global symmetries is the symmetric rank-2 stress-energy tensor of the vortex lattice:
\ie
&J_{it}\equiv p_i=n_va_i+n_v b_0\epsilon_{ij} u_j~,
\\
&J_{ij}=n_va_t\delta_{ij}-2\mu u_{ij}-\lambda u_{kk}\delta_{ij}~,
\fe
with $p_i$ denoting the momentum density.
It obeys the current conservation equation:
\ie
\partial_t J_{it}=\partial_j J_{ij}~,
\fe
that follows from the Euler-Lagrange equation \eqref{eq:EOM2} of $u_i$.
The conserved charges are  the momentum $P_i$ and the orbital angular momentum $L$ of the vortex lattice:
\ie\label{eq:momentum_charge}
P_i=\int d^2x\, p_{i}~,\quad L=\int d^2x\, \epsilon_{ij}x_ip_{j}~.
\fe

The conjugate momenta of $u_i$ and $a_i$ are $\pi_i=p_i$ and $\Pi_i=n_v u_i$ respectively. Upon quantization, 
\ie
{[\pi_i(x),u_j(x')]}&=i\delta_{ij}{\delta^2(x-x')}~,\\
{[\pi_i(x),a_j(x')]}&=0~,\\
{[\Pi_i(x),u_j(x')]}&=0~,\\
{[\Pi_i(x),a_j(x')]}&=i\delta_{ij}{\delta^2(x-x')}~.
\fe
These commutation relations reduce to more basic ones:
\ie\label{eq:commutation_relations}
{[u_i(x),u_j(x')]}&=0~,
\\
{[u_i(x),a_j(x')]}&=\frac{i}{n_v}\delta_{ij}\delta^2(x-x')~,
\\
{[a_i(x),a_j(x')]}&=-\frac{i b_0}{n_v}\epsilon_{ij}{\delta^2(x-x')}~.
\fe
Using these commutation relations, we obtain the following algebra of the momentum density of the vortex lattice:
\ie
{[p_i(x),p_j(x')]}= i n_vb_0\epsilon_{ij}\delta^2(x-x')~,
\fe
where $b_0=2\pi n_p$. The momentum operator thus obeys
\ie\label{eq:translation_algebra}
[P_i,P_j]= iN_v b_0\epsilon_{ij}~,
\fe
where $N_v$ is the total number of vortices.  This non-trivial algebra is a consequence of the Berry phase term in \eqref{eq:micro_Lagrangian}.
It is similar to the algebra obeyed by the magnetic translation operator $\hat P_i=-i\partial_i-A_i$ in a background magnetic field $B=\epsilon^{ij}\partial_i A_j$:
\ie
{[\hat{P}_i,\hat{P}_j]}=iB\epsilon_{ij}~.
\fe
This is not a coincidence, as the vortices are coupled to the vector gauge field $a_\mu$ of the superfluid and experience an effective magnetic field $b_0$, proportional to boson number density.

The algebra \eqref{eq:translation_algebra} has a c-number on the right-hand side, which can also be expressed as $iB N_0 \epsilon_{ij}$ with $N_0$ the total boson number in equilibrium. Physically, this algebra captures  a mixed anomaly between the orthogonal lattice translation symmetries when the boson number is nonzero in equilibrium. It is to be contrasted with the noncommutative algebra discussed in \cite{PhysRevResearch.6.L012040} where the boson number on the right-hand side is promoted to an operator. Such noncommutative algebra implies that the full symmetry group is an extension of the translation symmetry by the $U(1)$ symmetry associated with boson number conservation. The algebra reduces to our algebra \eqref{eq:translation_algebra} by approximating the boson number operator by its expectation value.

\subsection{Multipole symmetry of the Lifshitz theory}
The Lifshitz theory \eqref{eq:phi_Lagrangian} exhibits a quadrupole symmetry that shifts
\ie\label{eq:quadrupole-sym}
\phi\rightarrow\phi+c+c_xB  y+\frac{1}{2}c_yB(y-\sqrt{3}x)+\frac{1}{2}\alpha Br^2.
\fe
The associated current is
\ie\label{eq:quadrupole-current}
J_t^\phi= \chi\partial_t \phi-\frac{b_0}{B}\delta j_t~,
\quad J_{ij}^\phi=\frac{2\mu}{B^2}D_{ij}\phi+\frac{b_0}{2\pi B}\epsilon_{ij}\partial_t\phi~,
\fe
where 
\be\label{jtvortex}
\delta j_t=\frac{1}{2\pi}\epsilon_{ij} D_{ij}\phi=\frac{1}{2\pi }\mathbf{\grad}\times(\mathbf{\grad}\phi)
\ee
is the density of vortex interstitials.
It obeys the following current conservation equation corresponding to the Euler-Lagrange equation for $\phi$:
\ie
\partial_t J_t^\phi+D_{ij}J_{ij}^\phi=0~.
\fe
Similar quadrupole symmetry has also featured in classical vortex systems in a two dimensional incompressible fluid \cite{doshi2021vortices}.

We can construct four conserved charges from this current. They are respectively the integral, the first-moment integral along $x$ and $y$ direction, and the second-moment integral of $J_t^\phi$:
\ie\label{eq:charges}
&Q=-\int d^2x\ J_t^\phi~,
\\
&P_i=\int d^2x\ B \epsilon_{ij} x_j J_t^\phi~,
\\
&L=-\int d^2x\ \frac{1}{2}B r^2 J_t^\phi~.
\fe

The monopole charge $Q$ generates the constant shift on $\phi\rightarrow\phi+c$. Through the operator correspondence \eqref{eq:op_map}, it is mapped to the charge $Q$ in \eqref{eq:particle_charge} that measures the fluctuation of the total boson number around the equilibrium.
Since $Q$ is an integer, the constant identification on $\phi$ is fixed to be $\phi\sim\phi+2\pi$. 

The dipole charge $P_i$ generates the linear shift on $\phi\rightarrow\phi + B\epsilon_{ij}  x_j$, which under the relation \eqref{eq:u_to_phi} acts on the displacement field $u_i$ as lattice translation $u_j\rightarrow u_j+\delta_{ij}$. It is therefore identified with the lattice momentum operator $P_i$ in \eqref{eq:momentum_charge}. Indeed, one can check that the two charges are related by the operator map \eqref{eq:op_map} and integration by parts. The conjugate momentum of $\phi$ is $J_t^\phi$ so, upon quantization, we have
\ie
{[J_t^\phi(x),\phi(x')]}=i\delta^2(x-x')~.
\fe
This, together with \rf{jtvortex}, leads to the following algebra of the current $J_t^\phi$:
\ie
{[J_t^\phi(x),J_t^\phi(x')]}=-\frac{ib_0}{2\pi B}\mathbf{\grad}\times[\mathbf{\grad}\delta^2(x-x')]~,
\fe
which can be understood as a consequence of the Berry term in \eqref{eq:phi_Lagrangian}.
Using integration by parts, the dipole charge $P_i$ reproduces the algebra of momentum operators in \eqref{eq:translation_algebra}.

Similarly, the quadrupole charge $L$ generates the quadratic shift on $\phi\rightarrow\phi -\frac{1}{2}B r^2$ that acts on $u_i$ as the lattice rotation $u_i\rightarrow u_i-\epsilon_{ij}x_j$ under the relation \eqref{eq:u_to_phi}. It is thus identified with the angular momentum  operator \eqref{eq:momentum_charge} of the vortex lattice. 

By Lifshitz duality encoded in \eqref{eq:op_map_tensor_gauge_theory}, the traceless symmetric tensor gauge theory is manifestly invariant under the quadrupole symmetry \eqref{eq:quadrupole-sym}. The corresponding currents \eqref{eq:quadrupole-current}, expressed in terms of $\d_t\phi$ and $D_{ij}\phi$, can be expressed in terms of the dual field strength $\hat{\mathcal{B}},\hat{\mathcal{E}}_{ij}$ and the displacement field $u_i$:
\ie
J_t^\phi&= \dfrac{1}{2\pi}(\hat{\mathcal{B}}-b_0\partial_i u_i)-\dfrac{b_0B}{8\pi^2\mu}\epsilon_{jk}\hat{\mathcal{E}}_{ik}~,
\\
J_{ij}^\phi&=\dfrac{1}{2\pi}\epsilon_{jk}\hat{\mathcal{E}}_{ik}+\frac{b_0}{4\pi^2\chi B}\epsilon_{ij}(\hat{\mathcal{B}}-b_0\partial_i u_i)~.
\fe
The operators charged under the quadrupole symmetry are then the displacement field $u_i$ and the magnetic monopole operator $e^{i\phi}$ of the tensor gauge field, where the latter has no local representation on the dual side.
 
\section{Topological crystalline defects}\label{sec:defect}

One common way to characterize phases and phase transitions (complementary to Landau's approach of breaking symmetry of a disordered symmetric state) is through a proliferation of topological defects that restore symmetry of the symmetry-broken state \cite{KT,HNmelting,Youngmelting}, as have been recently applied to quantum melting of crystals and its descendant states, particularly utilizing fracton-elasticity dualities~\cite{Pretko_2018,PhysRevLett.121.235301,PhysRevB.100.134113,RHvectorPRL2020,RevModPhys.96.011001,Zhai_2021,PhysRevLett.125.267601,PhysRevLett.121.235301,PhysRevB.100.134113,Kumar_2019}.

A vortex lattice disorders similarly by proliferating topological defects that we analyze below, as was first implemented in a 3d classical vortex lattice in Ref.~\cite{MarchettiLRSS}
and recently in a 2d quantum vortex crystal in Ref.~\cite{Nguyen-Gromov-Moroz}. The distinction from a crystal of bosonic atoms is that in a vortex lattice, time reversal symmetry is explicitly broken and vortices are nontrivially coupled to superfluidity (superflow) encoded in the dual $U(1)$ gauge theory, as detailed in Sec.~\ref{rotatedSFvortexlattice}. As we have seen, important consequences of this are the incompressibility constraint encoded in the transversality of the phonon displacement $\pt_i u_i = 0$ and a nontrivial Berry phase of the Lifshitz theory \rf{eq:phi_Lagrangian} and its dual that we derived in \rf{eq:Lagrangian-tensor-gauge-theory-2}.

There are three types of topological crystalline defects exhibited by a vortex lattice: disclinations, dislocations and vortex vacancies/interstitials \footnote{For topological crystalline defects, we are referring to defects associated with field configurations with non-trivial topology. This should not be confused with the symmetry operators/defects that implement the crystalline symmetries.}.
Below we discuss and formulate these vortex lattice topological defects in terms of Lifshitz theory and its dual traceless symmetric tensor gauge theory, and use them to summarize possible descendant phases and discuss their properties, leaving a detailed analysis of corresponding Higgs transitions to future research.

To this end we note that there are a number of consistent disordering transition routes of a fully ordered commensurate superfluid vortex crystal. However, these are constrained by the interrelation of the associated topological defects, most importantly that dislocations and vacancies/interstitials are dipoles and quadrupoles of the elementary disclination defects, respectively. This is encoded in e.g., dislocation ($\psi^\dagger_b$) -- disclination ($\psi^\dagger_s$) coupling, $\psi^\dagger_s(\xv-{\bf d}/2) \psi_s(\xv+{\bf d}/2)\psi_b(\xv)$, that demands that when the latter condenses, so does, necessarily the former. As such, phases in which some  defects have proliferated and condensed, but their multipoles  have not, are inconsistent and thus physically impossible~\cite{PhysRevLett.121.235301, Zhai_2021, Kumar_2019}. With this we now construct these defects in Lifshitz theory and its gauge-dual representations, as summarized in  Table~\ref{tab:defect}, and discuss the corresponding phases that will generically appear as descendants of a vortex crystal.

\begin{table*}
\renewcommand{\arraystretch}{2.4}
\begin{tabular}{|c|c|c|}
\hline
Vortex lattice & Lifshitz theory & \quad Traceless symmetric tensor gauge theory\quad\quad
\\
\hline
Disclination & $\phi(\mathbf{x})=-\dfrac{B}{12}\vartheta r^2$ & $\exp\Big[\dfrac{i B}{6}\int \hat{\mathcal{A}}_t\, dt\Big]$
\\
\ \renewcommand{\arraystretch}{1}\begin{tabular}{@{}c@{}}Dislocation \\ with Burgers vector $\mathbf{b}$\end{tabular}\ \ & \ {$\phi(\mathbf{x}) = \dfrac{B}{4\pi}(\mathbf{b}\cdot \mathbf{x}+2\vartheta \mathbf{b}\times\mathbf{x})$}\ \ &\ $\exp\Big[-\dfrac{i B}{2\pi}\int \mathbf{b}\times\mathbf{\grad}\hat{\mathcal{A}}_t\, dt\Big]$\
\\
Vortex interstitial & $\phi(\mathbf{x})=\vartheta$ & \ \ $\exp\Big[-\dfrac{i }{2}\int \mathbf{\grad}^2 \hat{\mathcal{A}}_t\, dt\Big]$\ \
\\
\hline
\end{tabular}
\caption{The topological crystalline defects of the vortex lattice in the Lifshitz theory description \eqref{eq:phi_Lagrangian} and in its dual traceless symmetric scalar-charge theory description \eqref{eq:Lagrangian-tensor-gauge-theory-2}. The phase $\phi(\xv)$ of the Lifshitz theory winds around the defect placed at the origin, as listed in the second column. In the tensor gauge theory, the defects are realized as Wilson defects, as listed in the third column. These defects are charged under the multipole one-form symmetry of \eqref{eq:winding-current}.}\label{tab:defect}
\end{table*}

\subsection{Vortex-lattice defects via winding defects in Lifshitz theory}

In the Lifshitz theory, the crystalline defects are realized as point singularities around which the phase $\phi(\xv)$ winds. For such defects at the origin, the most general winding of $\phi(\xv)$ is fixed by the identification \eqref{eq:phi_identification}, \eqref{eq:2pi-phi} and is given by
\ie\label{eq:winding_number}
&\,\phi(\vartheta+2\pi,r)
\\
=&\,\phi(\vartheta,r)+2\pi \mathbb{Z}+yBl\mathbb{Z}+\frac{1}{2}(y-\sqrt{3}x)Bl\mathbb{Z}+\frac{2\pi}{12}Br^2\mathbb{Z}~,
\fe
where $(\vartheta,r)$ is the polar coordinate related to the Cartesian coordinate by $(x,y)=(r\cos\vartheta,r\sin\vartheta)$. Below, we relate crystalline defects of the vortex lattice to the winding defects in phase $\phi$ of the Lifshitz theory.

\subsubsection{Disclination}

A disclination defect is a singularity of the local lattice bond angle $\theta=\frac{1}{2}{\epsilon_{ij}\d_i u_j}$, which, utilizing \eqref{eq:u_to_phi} is given by 
\ie\label{eq:bond-angle-phi}
\theta=-\frac{1}{2B}\d^2\phi~,
\fe
in the Lifshitz theory under the operator map \eqref{eq:op_map}.
As the vortex lattice is a triangular lattice, the bond angle $\theta$ winds by ${2\pi}/{6}$ around an elementary disclination defect, leading to the following singularity equation for $\phi$:
\ie\label{eq:diclination}
-\frac{1}{2B}\epsilon_{ij} \d_i\d_j \d^2\phi=\epsilon_{ij} \d_i\d_j \theta=\frac{2\pi}{6}\delta^2(x)~.
\fe
Solving it gives
\ie\label{eq:disclination_phi}
\phi=-\frac{B}{12}\vartheta r^2~,
\fe
which corresponds to the quadratic winding in \eqref{eq:winding_number}, with  $\vartheta$ the polar angle singular at the defect location $r=0$.
The energy of a disclination: 
\ie
E_{disc}=\int_a^L rdr\int_0^{2\pi}d\theta  \frac{\mu}{ B^2}\left(D_{ij}\phi\right)^2=\frac{2\pi\mu}{144} L^2 ~,
\fe
diverges quadratically with the system size $L$, consistent with the studies of 2d classical crystals~\cite{HNmelting,Youngmelting}.

\subsubsection{Dislocation}

A dislocation defect is characterized by its Burgers vector $\mathbf{b}$ -- the shift by a lattice vector when going around the defect, possible because of the compactness of $\bf u(\xv)$. It can be decomposed into a dipole of the elementary disclinations separated by the vector $\frac{6}{2\pi}
\hat{\mathbf{z}}\times\mathbf{b}$. As a result, the $\phi$ configuration around it is related to the one in \eqref{eq:disclination_phi} around a disclination by the action of a differential operator $-\frac{6}{2\pi}\mathbf{b}\times\mathbf{\grad}$, which gives
\ie\label{eq:phi_dislocation}
\phi = \frac{B}{4\pi}\epsilon_{ij}b_i\d_j(\vartheta r^2)
=\frac{B}{4\pi}b_i(x_i+2\vartheta\epsilon_{ij} x_j)
~.
\fe
As $\vartheta$ increases by $2\pi$, $\phi(\xv)$ winds around by
\ie
\phi(\vartheta+2\pi)&=\phi(\vartheta)+B \epsilon_{ij} b_ix_j
\\
&=\phi(\vartheta)+ yB l\mathbb{Z}+\frac{1}{2}(y-\sqrt{3}x)B l\mathbb{Z}~,
\fe
where we use the fact that the Burgers vector are integer linear combinations of the triangular lattice vectors:
\ie
\mathbf{b}=\left\{\hat{\mathbf{ x}}\mathbb{Z}+\frac{1}{2}(\hat{\mathbf{ x}}+\sqrt{3}\hat{\mathbf{ y}})\mathbb{Z}\right\}l~.
\fe
This $\phi$ winding around a dislocation corresponds to the linear winding in the most general identification in \eqref{eq:winding_number}.  
The energy of a dislocation is given by
\ie
E_{v}=\int_a^L rdr\int_0^{2\pi}d\theta  \frac{\mu}{ B^2}\left(D_{ij}\phi\right)^2=\frac{\mu}{2\pi}\mathbf{b}^2\log\left(\frac{L}{a}\right)~,
\fe
diverging logarithmically with the system size $L$ as in a conventional 2d classical crystal~\cite{HNmelting,Youngmelting}.

\subsubsection{Vacancy and interstitial}

A vacancy/interstitial of the vortex lattice can be created by removing/inserting a vortex into the lattice. In the effective field theory \eqref{eq:phi_Lagrangian}, a vortex is represented by a Wilson line $W=\exp(i\int a_t\, dt)$ of the superfluid gauge field $a_\mu$. Consequently, a vortex vacancy and interstitial are described by $W^\dagger$ and $W$, respectively.  Without loss of generality, we will focus on the interstitial defect in the discussion below.

Inserting a vortex interstitial modifies the Euler-Lagrangian equation of $a_t$, resulting in a singularity in the incompressibility condition, modifying \eqref{eq:EOM1} to be
\ie\label{eq:incompressible_modify}
n_v\d_i u_i=\frac{1}{2\pi}\epsilon_{ij}\d_i\d_j \phi=\delta^2(x)~.
\fe
Here we employed the operator map in \eqref{eq:op_map} to replace $u_i$ with $\phi$.
Solving this equation gives the $\phi(\xv)$ configuration around the interstitial defect:
\ie\label{eq:phi_vortex}
\phi(r,\vartheta) =\vartheta~,
\fe
which corresponds to the constant winding component in \eqref{eq:winding_number}.
This $\phi$ configuration is related to the one around a disclination defect in \eqref{eq:disclination_phi}  by the action of the differential operator $-\frac{3}{B}\partial^2$.
This relationship arises from the fact that an interstitial defect can be decomposed into a group of disclination defects.  
The energy of a vacancy/interstitial is given by
\ie
E_{v}=\int_a^L rdr\int_0^{2\pi}d\theta  \frac{\mu}{ B^2}\left(D_{ij}\phi\right)^2=\frac{2\pi\mu}{B^2a^2}~,
\fe
finite in the infrared limit
but, as expected, depending sensitively on the lattice cutoff $a$.

\subsection{Multipole one-form global symmetry and defect mobility}

An interesting characteristic of these crystalline defects is that they could have restricted mobility, similar to fractons. To systematically organize them and study their mobility, we utilize the generalized global symmetries that they are charged under. 

In the Lifshitz theory, the relevant symmetry is the winding symmetry with the winding conserved current:
\ie\label{eq:winding-current}
J_t^{ij}= -\frac{1}{2\pi B}D_{ij}\phi~,\quad J=-\frac{1}{2\pi B}\partial_t \phi~,
\fe
where $J_t^{ij}$ is a traceless symmetric tensor. In the absence of topological defects, this winding current obeys the conservation equation:
\ie\label{eq:winding-conservation-eq}
\partial_t J_t^{ij}=D_{ij}J~,
\fe
and a differential condition relating different temporal components of the current:
\ie\label{eq:winding_conservation}
\epsilon_{jk}D_{ij}J_t^{ik}=0~.
\fe
These equations, \rf{eq:winding-conservation-eq} and \rf{eq:winding_conservation}, are the Lifshitz theory analogs akin to the spatial and temporal components of \rf{vorticity} for vortices in the superfluid. Here too, defects violate these homogeneous continuity equations, appearing as sources in it. In particular, a static defect generates only sources to the differential condition \eqref{eq:winding_conservation}.

Using the winding current, we can construct several conserved charges, e.g.\ $Q^{xy}_i=\int J_t^{xy} dx_i$, that act on the winding operators and the winding states of $\phi$. Their conservation follows from the current conservation equation \eqref{eq:winding-conservation-eq}. 
Notably, these conserved charges act only on operators inserted at a fixed time but not on static defects extending in the time direction. In the language of~\cite{Gorantla:2022eem}, these charges generate a space-like symmetry, while the defects are charged under a time-like symmetry. As defects are sources to the differential conditions \eqref{eq:winding_conservation},  the time-like charges can be built from the latter.

As a warm-up, we recall the simplest example of space-like and time-like symmetries is the $U(1)$ electric one-form symmetry in Maxwell's electrodynamics, generated by a conserved two-form current:
\ie
J_{\mu\nu}=\frac{1}{g^2}F_{\mu\nu}~,
\fe
obeying the current conservation equation (Faraday's law):
\ie\label{eq:conserve-1-form}
\partial_t J_{ti}=\partial_j J_{ji}~,
\fe
and a differential condition relating different temporal components of the current (Gauss's law):
\ie\label{eq:diff-condition-1-form}
\partial_i J_{it}=0~.
\fe
The conservation and differential condition can be summarized concisely as $\partial^\mu J_{\mu\nu}=0$.
The $U(1)$ one-form symmetry comprises a $U(1)$ space-like symmetry that acts on Wilson loop operators inserted at a fixed time and a $U(1)$ time-like symmetry that acts on Wilson line defects oriented in the time direction. The time-like charge for the time-like symmetry is an integral of the differential condition \eqref{eq:diff-condition-1-form} over an open volume $\mathcal{V}$:
\ie
Q&\,=\int_\mathcal{V} \partial_i J_{it}\, dx\wedge dy\wedge dz
\\
&\,=\oint_{\Sigma} \frac{1}{2}\epsilon^{ijk} J_{it}\,dx_j\wedge dx_k
\\
&\,=\oint_\Sigma \mathbf{E}\cdot d\mathbf{S}~,
\fe
which can be expressed as the Gauss's law operator localized on the boundary surface $\Sigma=\partial\mathcal{V}$. A Wilson defect induces a source to the differential condition \eqref{eq:diff-condition-1-form} and thus carries charge under the time-like charges or equivalently the Gauss's law operators surrounding it. This is essentially the content of Gauss's law. The time-like charge is conserved in time if there is no Wilson line operators crossing it:
\ie
\partial_tQ=\int_\mathcal{V} \partial_i \partial_tJ_{it}\, dv=\int_\mathcal{V} \partial_i \partial_jJ_{ij}\, dv=0
\fe
where $dv=dx\wedge dy\wedge dz$ denotes the volume form. In the second equality, we use the current conservation equation \eqref{eq:conserve-1-form} while in the last equality, we use the fact that $J_{ij}$ is antisymmetric and single-valued, namely there are defects so that derivatives on it commute. 

In relativistic systems, the space-like and time-like symmetries share the same symmetry group and together form a higher-form symmetry since space and time are on equal footings, but in non-relativistic systems, they are generically distinct (see the examples discussed in \cite{Gorantla:2022eem}). 

After the warm-up, we are now ready to explore the time-like winding  symmetry in the Lifshitz theory. The time-like symmetry is a multipole symmetry, with in total, three types of time-like winding charges---monopole, dipole and quadrupole charges---paralleling our classification  of winding defects in the previous subsection, and in particular the three---constant, linear and quadratic--- components of \rf{eq:winding_number}.

The monopole charge $Q$ is the integral of the differential condition \eqref{eq:winding_conservation} over an open surface area $\Sigma$:
\ie
Q=&\,\int_{\Sigma} \epsilon_{jk}D_{ij} J_t^{ik} \, dx\wedge dy
\\
=&\,\oint_{\gamma}\d_i J_t^{ij}dx_j
\\
=&-\oint_\gamma \frac{1}{4\pi B}d(\d^2\phi)=\oint_{\gamma}\frac{1}{2\pi}d\theta~,
\fe
which can be expressed as an operator localized on the boundary curve $\gamma=\partial\Sigma$. In the last two equalities, we used the explicit form of the winding current \eqref{eq:winding-current} and the relation \eqref{eq:bond-angle-phi} between the bond angle $\theta$ and the Lifshitz scalar field $\phi$. This reveals the physical meaning of the monopole charge: It  measures the quadratic winding of $\phi$ in \eqref{eq:winding_number} or equivalently the bond angle $\theta$ winding around the curve $\gamma$, nonzero for an enclosed disclination defect.

The dipole charge $Q_n$ is the first-moment integral of the differential condition \eqref{eq:winding_conservation} 
over an open area $\Sigma$:
\begin{widetext}
\ie\label{eq:dipole-timelike-charge}
Q_n=&\,\int_\Sigma 2\pi \epsilon_{nm}x_m\times \epsilon_{jk}D_{ij}J^{ik}_t\,dx\wedge dy
\\
=&\,\oint_{\gamma}2\pi\epsilon_{n\bar{n}} \big[x_{\bar{n}}\big(\partial_n J_t^{xy}+2\partial_{\bar{n} }J_t^{\bar{n}\bar{n}}\big)dx_{\bar{n}}+\big(x_{\bar{n}}\partial_{\bar{n}}J_t^{xy}- J_t^{xy}\big)dx_n\big]~,
\\
=&\,\oint_{\gamma}\frac{1}{B} \epsilon_{n\bar{n}} d\left(\partial_{\bar{n}}\phi-x_{\bar{n}}\partial_{\bar{n}}^2\phi\right)=\oint_{\gamma} d\left(u_n+\theta \epsilon_{n\bar n}x_{\bar{n}}\right)~.
\fe
where in the last line, the indices are not summed over; instead, $\bar n$ denotes the index distinct from $n$.
It can be expressed as an operator localized on the boundary $\partial\Sigma=\gamma$, which measures the linear winding of $\phi$ in \eqref{eq:winding_number}, corresponding to a Burgers displacement $\bf b$ associated with an enclosed dislocation defect. The dipole charge is not simply an integral of $d(\partial_{\bar n}\phi)$; rather, it has an additional term $-d(x_{\bar n}\partial^2_{\bar n}\phi)$ that removes the contributions from the quadratic winding. In the last equality, we expressed the dipole charge in terms of the displacement field $u_n$ and the bond angle $\theta$ using \eqref{eq:u_to_phi} and  \eqref{eq:bond-angle-phi}. This allows us to interpret the linear-in-$x$ winding of $\phi$  physically as measuring the variation of the displacement field $u_n$ around the curve $\gamma$ (dislocation), modulo the contribution $\theta \epsilon_{n\bar n}x_{\bar{n}}$ from the winding of the bond angle $\theta$ (disclination).

The quadratic charge $\mathcal{Q}$ is the second-moment integral of the differential condition \eqref{eq:winding_conservation} 
over an open area $\Sigma$:
\ie\label{eq:time_like_charge_quadratic}
\mathcal{Q}&\,=\int_\Sigma -\frac{1}{2}Br^2\times\epsilon_{jk}D_{ij}J^{ik}_t \,dx\wedge dy
\\
&\,=\oint_{\gamma}\frac{1}{2}B \left[\left(2yJ_t^{xy}-r^2\partial_y J_t^{xy}-2x^2\partial_x J_t^{xx}\right)dx+\left(2xJ_t^{xy}-r^2\partial_x J_t^{xy}-2y^2\partial_y J_t^{yy}\right)dy\right]~,
\\
&\,=\oint_{\gamma}\frac{1}{4\pi} d(x^2\partial_x^2\phi-2x\partial_x\phi+y^2\partial_y^2\phi-2y\partial_y\phi+2\phi)=\int_\Sigma \delta j_t-n_v\oint_{\gamma}d\Big(\epsilon_{ij} u_ix_j+\frac{1}{2}r^2\theta\Big)~.
\fe
\end{widetext}
It can be expressed as an operator localized on the boundary $\partial\Sigma=\gamma$, which measures only the constant winding of $\phi$ in \eqref{eq:winding_number}. Physically, the quadrupole charge can be interpreted as the number of vortices enclosed by the curve $\gamma$ modulo the contributions from the underlying vortex lattice and the change of lattice area $\epsilon_{ij} u_ix_j+\frac{1}{2}r^2\theta$ due to dislocations and disclinations, namely it measures vortex vacancies and interstitial defects added to the underlying vortex lattice.

We can systematically organize the winding defects or equivalently the crystalline defects according to their time-like winding charges. Let us first consider only defects placed at the origin. Then, (i) defects charged under the monopole charge $Q$ are the disclination defects with quantized $Q\in\frac{1}{6}\mathbb{Z}$ measuring the winding of the bond angle $\theta$, (ii) defects charged under the dipole charges $Q_n$ are the dislocation defects with $Q_n=b_n$ measuring the Burgers vector of the dislocation defects $\mathbf{b}$, and (iii) defects charged under the quadrupole charge $\mathcal{Q}$ are vortex vacancies/interstitials with quantized $\mathcal{Q}\in\mathbb{Z}$ measuring the constant winding of $\phi$, or equivalently the constant winding of the superfluid phase $\varphi$ over and above the underlying winding due to the vortex lattice.

For defects away from the origin, the ones charged under the dipole and quadrupole time-like symmetry are no longer just the dislocations and vortex vacancies/interstitials respectively. Comparing the singularity equation \eqref{eq:diclination} with the differential condition \eqref{eq:winding_conservation},  we learn that a disclination defect generates a $\delta$-function source to the differential condition:
\ie\label{eq:disclination-differential-condition}
\epsilon_{jk}D_{ij} J_t^{ik}=\frac{1}{6}\delta^2(\mathbf{x}-{\mathbf{x}'})~,
\fe
where $\mathbf{x}'$ is the position of the disclination defect. Inserting \rf{eq:disclination-differential-condition} into \rf{eq:dipole-timelike-charge} and \rf{eq:time_like_charge_quadratic} we see that such disclination carries both a dipole charge $Q_n=\frac{2\pi}{6} \epsilon_{nm} {x}'_m$ and a quadrupole charge $\mathcal{Q}=\frac{1}{12}Br'^2$, respectively. Conservation of these charges requires that $\mathbf{x}'$ is time independent, and thus a disclination is completely immobile. Physically, it corresponds to the fact that moving a disclination creates dislocations~\cite{RHvectorPRL2020}. Similarly, a dislocation carries a quadrupole charge $\mathcal{Q}=\frac{1}{2\pi} B\mathbf{b}\times{\mathbf{x}'}$ which freezes its position transverse to its Burgers vector $\bf b$, forbidding its ``climb''. We thereby recover the glide-only constraint on a dislocation motion. Physically, this corresponds to a creation of vortex vacancies and interstitials in the climb process, that is forbidden by vortex conservation, as explicitly demonstrated in Refs.~\cite{Pretko_2018,PhysRevLett.121.235301,PhysRevB.100.134113,RHvectorPRL2020}.

\subsection{Vortex-lattice defects via Wilson defects in tensor gauge theory}
The topological crystalline defects of the vortex lattice are realized as Wilson defects in the traceless symmetric tensor gauge theory \eqref{eq:Lagrangian-tensor-gauge-theory-2}. A systematic way to match these defects is to utilize the generalized global symmetry that acts on them. In the Lifshitz theory, the relevant symmetry is the winding symmetry whose conserved current under the operator correspondence \eqref{eq:op_map_tensor_gauge_theory} is mapped to 
\ie\label{eq:electric-current}
J_t^{ij}= -\dfrac{B}{8\pi^2\mu}\epsilon_{jk}\hat{\mathcal{E}}_{ik}~,\quad J=-\dfrac{1}{4\pi^2 B\chi}(\hat{\mathcal{B}}-b_0\partial_i u_i)~.
\fe
The differential condition \eqref{eq:winding_conservation} of this current
now is a consequence of the Euler-Lagrange equation of $\hat{\mathcal{A}}_t$ in the traceless symmetric tensor gauge theory, namely the electric Gauss's law for $\hat{\mathcal{E}}_{ik}$. The sources to the differential conditions are the Wilson lines built out of $\hat{\mathcal{A}}_t$. Suppose we insert such a Wilson line $\exp(iC\int \hat{\mathcal{A}}_t\, dt)$ with an undetermined coefficient $C$ at the origin. The Euler-Lagrangian equation or relatedly the differential condition is modified into
\ie\label{eq:modified-EOM}
\epsilon_{jk}D_{ij}J_t^{ik}=\frac{B}{8\pi^2\mu}D_{ij}\hat{\mathcal{E}}_{ij}=\frac{C}{B}\delta^2(x)~,
\fe
with Gauss's law encoding disclinations as the (dual) electric charges.
Comparing it with the differential condition \eqref{eq:disclination-differential-condition} around the elementary disclination defect, we learn that an elementary disclination defect is represented by the Wilson line:
\ie\label{eq:basic-wilson-line}
\text{vortex-disclination:}\quad\exp\left[\frac{i B}{6}\int \hat{\mathcal{A}}_t\, dt\right]~.
\fe
The exponent of the Wilson line is dimensionless despite the appearance of the effective background magnetic field $B$ in it. It is because the gauge parameter $\hat\lambda$ has mass dimension $-2$. Since this Wilson line must be the minimal properly quantized Wilson line, the gauge parameter $\hat\lambda$ has the following identification:
\ie
\hat\lambda\sim \hat\lambda +\frac{12\pi}{B}\mathbb{Z}~.
\fe
As dislocations and vortex interstitials are composed of disclinations, we can infer their representations by acting differential operators on the exponent of Wilson line representation \eqref{eq:basic-wilson-line} of disclinations. With this, we deduce that the dislocation with Burgers vector $\mathbf{b}$:
\be\label{dislcoationBurgers}
\mathbf{b}=\left\{\hat{\mathbf{ x}}\mathbb{Z}+\frac{1}{2}(\hat{\mathbf{ x}}+\sqrt{3}\hat{\mathbf{ y}})\mathbb{Z}\right\}l,
\ee
and the vortex interstitial are respectively represented by the following properly quantized Wilson defects:
\begin{alignat}{2}
\label{dislocWilson}\text{vortex-dislocation:}&\quad&&\exp\left[-\frac{i B}{2\pi}\int \mathbf{b}\times\mathbf{\grad}\hat{\mathcal{A}}_t\, dt\right]~,
\\
\label{vacanciesWilson}\text{vortex-interstitial:}&\quad&&\exp\left[-\frac{i }{2}\int \mathbf{\grad}^2 \hat{\mathcal{A}}_t\, dt\right]~.
\end{alignat}
The quantization of these defects demands that the gauge parameter $\hat\lambda$ obey the following additional identifications:
\ie
\hat\lambda\sim\hat\lambda+\frac{8\pi^2}{\sqrt{3}Bl}[x\mathbb{Z}+\frac{1}{2}(x+\sqrt{3}y)\mathbb{Z}]+ \pi(x^2+y^2)\mathbb{Z}~.
\fe

The mobility of these defects is reflected in whether there exist gauge-invariant Wilson defects that can be deformed in spatial directions. If such defects exist, they represent the worldlines of the mobile defects. In the case of disclinations and dislocations, such gauge-invariant defects do {\it not} exist, so they are completely immobile, consistent with our earlier conclusion in the previous subsection and with general considerations based on fracton-elasticity duality \cite{Pretko_2018,PhysRevLett.121.235301,PhysRevB.100.134113,RHvectorPRL2020}. On the other hand, a vortex interstitial can move freely, and its motion is represented by the gauge-invariant Wilson defect:
\ie\label{eq:vortex-Wilson-line}
\exp\left[-i\int_\gamma\left(\frac{1}{2}\partial^2 \hat{\mathcal{A}}_t\, dt+\partial_i \hat{\mathcal{A}}_{ij}\,dx_j\right)\right]~,
\fe
where $\gamma$ is the worldline of the vortex interstitial. The Wilson defect is invariant under the gauge transformation \eqref{eq:traceless-gauge-field-maintext}, as the phase $\Phi$ generated by the transformation:
\ie
\Phi=\int_\gamma\left(\frac{1}{2}\partial^2 \partial_t\hat{\lambda}\, dt+\partial_i D_{ij}\hat\lambda\,dx_j\right)=\int_\gamma \frac{1}{2}d(\partial^2 \hat{\lambda}),
\fe
is the integral of an exact form and thereby vanishes.

The Wilson defect descriptions of the topological crystalline defects allow us to compute the potentials between them. Consider a disclination defect sitting at the origin. It modifies the equation of motion of $\mathcal{A}_t$, $\mathcal{A}_{ij}$ and $u_i$ to
\ie
-\frac{B^2}{8\pi^2\mu}D_{ij}\hat{\mathcal{E}}_{ij}+\frac{B}{6}\delta^2(x)&=0~,
\\
-\frac{B^2}{4\pi^2\mu}\partial_t\hat{\mathcal{E}}_{ij}-\frac{1}{2\pi^2\chi}\epsilon_{ik}D_{jk}(\hat{\mathcal{B}}-b_0\partial_iu_i)&=0~,
\\
-\frac{b_0}{4\pi^2\chi}\partial_i(\hat{\mathcal{B}}-b_0\partial_iu_i)-\frac{b_0 B}{2\pi}\epsilon_{ij}\partial_t u_j&=0~.
\fe
Solving them gives the solution:
\ie
\mathcal{A}_t=-\frac{\pi\mu}{3B}r^2(\log r-1)~,\quad\mathcal{A}_{ij}=0~,\quad u_i=0~.
\fe
The potential between a disclination and an anti-disclination is then given by 
\ie
V_d(r)=-\frac{B}{6}\mathcal{A}_t=\frac{\pi\mu}{18}r^2(\log r-1)~,
\fe
which describes a confining potential. Since the dislocations and vacancies/interstitials are compositions of disclinations, it is straightforward to derive their potentials. For example, the potential between a vacancy and interstitial is given by
\ie
V_v(r)=\left(\frac{3}{B}\mathbf{\grad}^2\right)^2V_d(r)=0~,
\fe
which vanishes. The potential gets corrected if the subdominant kinetic and electric energies were included in the effective theory \eqref{eq:micro_Lagrangian}. 

\section{Vortex phases and phase transitions}
\label{sec:phase}

We now turn to a discussion of phases whose general structure directly follows from various schemes of proliferation of topological defects:\ vacancies/interstitials, dislocations and disclinations.

\subsection{Vortex-supersolid} 
Per our comments above on generic unbinding the highest multipole defects first, we consider proliferating disclination quadrupoles, i.e., vacancies and interstitials, as discussed above, corresponding to vortices in $\phi$.  In the dual traceless symmetric tensor gauge theory picture, this condensation corresponds to Higgs transition of the vortex crystal to the ``vortex-supersolid''~\cite{Nguyen-Gromov-Moroz} (a novel, non-superfluid vortex lattice phase without ODLRO of the underlying bosons, that has been previously extensively studied in the context of type-II superconductors~\cite{FreyFisherNelsonSS,MarchettiLRSS}).
These quadrupole ``charges'' couple to an ordinary Wilson line $W=\exp(i\int_\gamma a_\mu dx^\mu)$ built from a composite vector gauge field in \eqref{eq:vortex-Wilson-line}:
\ie
a_\mu=(a_t,a_i)_\mu=-\Big(\frac{1}{2}\partial^2 \hat{\mathcal{A}}_t,\partial_j \hat{\mathcal{A}}_{ji}\Big)_\mu~.
\fe
The fluctuations of the quadrupole charges can be represented by a complex field $\Psi$, with the gauge symmetry:
\ie
\Psi\rightarrow e^{\frac{i}{2}\partial^2\alpha}\Psi ~.
\fe
Then the vortex-crystal-to-vortex-supersolid melting transition is captured by the following quantum Landau-Ginzburg theory:
\ie
\mathcal{L}=\,&\frac{B^2}{16\pi^2\mu}\hat{\mathcal E}_{ij}^2-\frac{1}{8\pi^2\chi}(\hat{\mathcal B}-b_0\partial_iu_i)^2
-\frac{b_0 B}{4\pi }\epsilon_{ij}  u_i \partial_t u_j
\\
&+K_t\big|\big(\partial_t-\frac{i}{2}\partial^2\hat{\mathcal{A}}_t\big)\Psi\big|^2+K\big|\big(\partial_i-i\partial_j\hat{\mathcal{A}}_{ji}\big)\Psi\big|^2
\\
&+m^2|\Psi|^2+g|\Psi|^4~.
\fe
When $m^2$ is positive, $\Psi$ is massive, and the theory is the vortex-lattice phase. When $m^2$ is negative, $\Psi$ condenses and develops a vacuum expectation value $\langle \Psi\rangle=v$. It can be parameterized as
\ie
\Psi=(v+\sigma)e^{i\mathcal{A}}~.
\fe
The phase $\mathcal{A}$ transforms under the gauge symmetry as a gauge field:
\ie
\mathcal{A}\rightarrow \mathcal{A}+\frac{1}{2}\partial^2\hat\lambda~.
\fe
Combining this gauge field $\mathcal{A}$ with the traceless symmetric gauge field $(\hat{\mathcal{A}}_t,\hat{\mathcal{A}}_{ij})$, we can define a \textit{traceful} symmetric tensor gauge field as
\ie
\mathcal{A}_t=\hat{\mathcal{A}}_t~,\quad \mathcal{A}_{ij}=\hat{\mathcal{A}}_{ij}+\delta_{ij}\mathcal{A}~,
\fe
which transforms under the gauge symmetry as
\ie
\mathcal{A}_t\rightarrow \mathcal{A}_t+\partial_t\hat\lambda~,\quad \mathcal{A}_{ij}\rightarrow \mathcal{A}_{ij} +\partial_i\partial_j\hat\lambda~.
\fe
The gauge-invariant field strength are
\ie
\mathcal{B}_i=\epsilon_{jk}\partial_j\mathcal{A}_{ki}~,\quad \mathcal{E}_{ij}=\partial_t \mathcal{A}_{ij}-\partial_i\partial_j \mathcal{A}_t~.
\fe
They are related to the field strength of $(\hat{\mathcal{A}}_t,\hat{\mathcal{A}}_{ij})$ by
\ie
\hat{\mathcal E}_{ij}= \mathcal{E}_{ij}-\frac{1}{2}\delta_{ij} \mathcal{E}_{kk}~, \quad \hat{\mathcal B}=-\partial_i \mathcal{B}_i~.
\fe
In terms of this new symmetric tensor gauge field, the vortex supersolid phase is described by the following Lagrangian:
\ie
\mathcal{L}=\,&\frac{B^2}{16\pi^2\mu}(\mathcal{E}_{ij}-\frac{1}{2}\delta_{ij} \mathcal{E}_{kk})^2+\frac{1}{4}K_t\mathcal{E}_{kk}^2+K\mathcal{B}_i^2
\\
&-\frac{1}{8\pi^2\chi}(\partial_i\mathcal{B}_i+b_0\partial_iu_i)^2
-\frac{b_0 B}{4\pi }\epsilon_{ij}  u_i \partial_t u_j~.
\fe
The Higgs transition thus leads to the dual {\it traceful} symmetric tensor gauge theory of the vortex-supersolid state (as opposed to a {\it traceless} one in \rf{eq:Lagrangian-tensor-gauge-theory-2}).  Interestingly, this dual tensor gauge theory resembles that of a time-reversal broken Wigner crystal (e.g., bosonic crystal in an effective magnetic field) \cite{PhysRevLett.121.235301}.

In this vortex-supersolid phase, the condensate of vortex vacancies and interstitials are encoded in the traces of $\mathcal{A}_{ij}$ and ${\mathcal{E}}_{ij}$. 
They break the $U(1)$ vortex conservation symmetry and thereby lift the glide-only constraint on the dislocation mobility, akin to its time-reversal symmetric analog of bosonic crystals~\cite{PhysRevLett.121.235301,PhysRevB.100.134113,RHvectorPRL2020,Kumar_2019}. 
The motions of these mobile vortex-dislocations are represented by the gauge-invariant Wilson defect of the symmetric tensor gauge field:
\ie
\exp\left[-\frac{i B}{2\pi}\int \epsilon_{ij}b_i(\d_j\mathcal{A}_t\, dt+\mathcal{A}_{jk} dx_k)\right]~.
\fe
Physically, vortex-dislocations become fully mobile in the presence of vortex condensate because the latter absorbs the vortex vacancies/interstitials created by the motions of dislocations.
Contrary to the vortex-dislocations, vortex-disclinations remain immobile in the vortex-supersolid phase~\cite{MarchettiLRSS}.

\subsection{Vortex-hexatic, -smectic and -nematic}
\indent{\it Vortex-hexatic}: 
Next, a simultaneous proliferation of all dislocations ${\bf b}_\alpha$ (dipoles) in \rf{dislcoationBurgers} melts the vortex-supersolid into vortex-hexatic, fully restoring the translational symmetry while retaining bond orientational order.  Like its vortex-supersolid parent, this state lacks ODLRO of bosons, i.e., it is not a superfluid.
Extending arguments from melting of time-reversal invariant crystals of bosons to here, we find that the corresponding hexatic vortex-liquid state is characterized by mobile disclinations. Physically, this is due to a condensate of dislocations now able to absorb dislocations produced by disclination motion, and characterized by a space component of the disclination Wilson loop in Table I. 

\indent{\it Vortex-smectic}: 
If in contrast to vortex-hexatic, only one of the three vortex-dislocations in \rf{dislcoationBurgers} unbinds, the 
condensation of such dislocation $\bf b$ melts the vortex-crystal into a vortex-smectic, characterized by a wavevector transverse to the condensed $\bf b$ and a vanishing shear modulus for shear along $\bf b$. The Higgs transition gaps out gauge fields associated with the phonon displacement along $\bf b$ and leads to restricted lineon mobility of the corresponding disclinations, immobile only along smectic layers. 

\indent{\it Vortex-nematic}:
Unbinding the remaining two elementary dislocations (they must unbind in pairs), leads to a non-superfluid vortex-nematic state with fully mobile disclinations, quite similar to a hexatic, but with $C_2$ symmetry. 

Both the hexatic and nematics can then quantum melt into an isotropic vortex liquid via a Higgs condensation transition of disclination defects.
We leave the detailed derivation and the resulting field theory to later studies.

\section{Conclusions}
In this paper, we formulated and studied an effective quantum field theory of a two-dimensional zero-temperature vortex lattice in a neutral rotating superfluid, utilizing a combination of particle-vortex, elasticity-fracton and Lifshitz-gauge dualities, augmented with
a Berry phase term that encodes vortex dynamics in the
presence of a superflow, demonstrating consistency between these different formulations.  We discussed a hidden multipole symmetry of the
vortex lattice, used its dual traceless symmetric tensor gauge theory to explore vortex lattice dynamics, characterized its topological vortex crystalline defects using a multipole one-form symmetry and generalized Wilson lines, uncovering and detailing their restricted mobility.  
This also allowed us to outline a number of novel descendant vortex phases separated by generalized-Higgs transitions, with detailed analysis left for future studies.

It would be interesting to generalize our vortex lattice duality to nonlinear Lifshitz theory. We also expect that the duality can be extended to higher-dimensional theories, e.g.\ the Lifshitz photon theory~\cite{PhysRevB.109.035135} in three dimensions, which is a gauge theory with Lifshitz scaling symmetry.

\noindent\textit{Note added}: We would like to draw the reader's attention  to the paper~\cite{Dung:2023} by Nguyen and Moroz, titled ``On quantum melting of superfluid vortex crystals: from Lifshitz scalar to dual gravity", that has some overlap with our work.
\acknowledgments
The authors thank Dam Thanh Son for insightful discussions, and Dung Nguyen and Sergej Moroz for discussion of their work. YHD was supported, in part, by the U.S.\ DOE Grant No.\ DE-FG02-13ER41958 and by the Simons Collaboration on Ultra-Quantum Matter, which is a grant from the Simons Foundation (651440, D.T.S.). YHD would also like to thank the Kavli Institute for Theoretical Physics, University of California, Santa Barbara, supported in part by the National Science Foundation under Grant No. NSF PHY-1748958, the Heising-Simons Foundation, and the Simons Foundation (216179, LB). HTL was supported in part by a Croucher fellowship from the Croucher Foundation, the Packard Foundation and the Center for Theoretical Physics at MIT. HTL would like to thank the Perimeter Institute for Theoretical Physics for their hospitality, when part of this research was carried out. Research at Perimeter Institute is supported in part by the Government of Canada through the Department of Innovation, Science and Economic Development
Canada and by the Province of Ontario through the Ministry of Colleges and Universities. LR was supported by The Simons Investigator Award from the Simons Foundation. LR and YHD also thank The Kavli Institute for Theoretical Physics for hospitality during Quantum Crystals and Quantum Magnetism workshops, when part of this research was carried out, supported by the National Science Foundation under Grant No. NSF PHY-1748958 and PHY-2309135.

\onecolumngrid

\appendix

\section{Duality of Lifshitz theory without Berry phase}\label{app:duality-no-berry}
In Sec.~\ref{sec:Lifshtiz-duality}, we study the duality of Lifshitz theory with a Berry phase presented in \eqref{eq:phi_Lagrangian}. In this appendix, we instead consider  Lifshitz theory without the Berry phase term and revisit the Lifshitz duality in the presence of higher-rank background probe fields~\cite{Du_2022}. 
\subsection{Coupling to the higher-rank background probe fields}\label{app:coupling-linear}
The Lagrangian is given by
\ie\label{eq:lifshitz-Lag-appedix}
\mathcal{L}= \frac{\chi}{2}\left(\partial_t \phi-\hat C_t\right)^2-\frac{K}{2}\left(D_{ij}\phi-\hat C_{ij}\right)^2~,
\fe
where $(\hat C_t, \hat C_{ij})$ is the traceless symmetric background gauge field with background gauge transformation:
\ie\label{eq:gauge-higher-rank}
\phi\sim \phi+\gamma~,\quad \hat C_t\sim  \hat C_t+\partial_t\hat \gamma~,\quad \hat C_{ij}\sim \hat C_{ij}+D_{ij}\hat\gamma~.
\fe

We dualize the theory \eqref{eq:lifshitz-Lag-appedix} by introducing Hubbard-Stratonovich fields $\hat \pi_t$ and $\hat \pi_{ij}$, which is symmetric and traceless, and rewrite the Lifshitz theory Lagrangian as 
\ie
\mathcal{L}=-\frac{1}{2\chi}\hat\pi_t^2+\frac{1}{2K}\hat\pi_{ij}^2
+\hat \pi_t(\partial_t\phi-\hat C_t)+ \hat \pi_{ij}(D_{ij}\phi-\hat C_{ij}) ~.
\fe
Integrating out $\phi$, that  appears linearly, gives the following constraint:
\ie
\partial_t\hat\pi_t-D_{ij}\hat \pi_{ij}=0~.
\fe
To solve the constraint, it is convenient to define dual magnetic and electric fields:
\ie
\hat{\mathcal{B}}\equiv 2\pi \hat\pi_t~,
\quad\hat{\mathcal{E}}_{ij}\equiv 2\pi \epsilon_{jk}\hat\pi_{ik}~,
\fe
in terms of which constraint transforms into Faraday-like law:
\ie\label{Farady-like-law-app}
\partial_t\hat{\mathcal{B}}-\epsilon_{ik}D_{jk}\hat{\mathcal{E}}_{ij}=0~.
\fe
Here $\hat{\mathcal{E}}_{ij}$ is a traceless symmetric tensor, following from the fact that $\hat \pi_{ij}$ is traceless symmetric. The constraint
can be solved by introducing a traceless symmetric tensor gauge field:
\ie\label{eq:gauge-higher-rank-dual}
\hat{\mathcal{A}}_t\sim \hat{\mathcal{A}}_t+\partial_t \hat\lambda~,
\quad \hat{\mathcal{A}}_{ij}\sim \hat{\mathcal{A}}_{ij}+D_{ij} \hat\lambda~,
\fe
and identifying $\hat{\mathcal{E}}_{ij}$, $\hat{\mathcal{B}}$ with the gauge-invariant field strengths of the tensor gauge field:
\ie\label{eq:gauge-invariant-field-strengths}
\hat{\mathcal{E}}_{ij}=\d_t  \hat{\mathcal{A}}_{ij} -  D_{ij}\hat{\mathcal{A}}_t~, \quad \hat{\mathcal{B}}=
\epsilon_{ik}D_{jk} \hat{\mathcal{A}}_{ij}~.
\fe
This then gives the dual tensor gauge theory Lagrangian:
\ie\label{eq:dual-tensor-theory}
\mathcal{L}=\frac{1}{8\pi^2K}\hat{\mathcal{E}}_{ij}^2-\frac{1}{8\pi^2\chi}\hat{\mathcal{B}}^2
-\frac{1}{2\pi}\hat C_t\hat{\mathcal{B}}+\frac{1}{2\pi}\epsilon_{jk}\hat C_{ij}\hat{\mathcal{E}}_{ik}~,
\fe
where the higher-rank background probe field $(\hat C_t,\hat C_{ij})$ couples to the tensor gauge theory via a generalized Chern-Simons term for the tensor gauge fields. The background gauge invariance of the coupling follows from the Faraday-like law in \eqref{Farady-like-law-app}. 

\subsection{Connections with linearized gravity}

The space components of the background traceless symmetric gauge field 
$\hat C_{ij}$ introduced in the previous section can be naturally interpreted in terms of a background linearized traceless symmetric metric $\hat H_{ij}$ with the relation~\cite{Du_2022}:
\ie
\hat H_{ij} = - \bar \ell^2 \left(\epsilon_{ik}\hat C_{jk}+ \epsilon_{jk} \hat C_{ik}\right)~, 
\fe
where $\bar \ell$ is a constant with dimension of length. The gauge transformation on $\hat C_{ij}$ \eqref{eq:gauge-higher-rank} implies transformation of the metric:
\ie
\hat H_{ij} \to \hat H_{ij} - \bar{\ell}^2 \left(\epsilon_{ik} \d_j \d_k + \epsilon_{jk} \d_i \d_k \right)\hat \gamma~,
\fe
which can be reformulated by linearized version of the transformation of the metric under area-preserving diffeomorphism:
\ie
\hat H_{ij} \to \hat H_{ij} - \d_i \xi_j -\d_j \xi_i~,
\fe
with the definition of $\xi_i=\bar\ell^2 \epsilon_{ik}\d_k \hat \gamma$. Similarly the traceless symmetric tensor gauge field $(\hat{\mathcal{A}}_t, \hat{\mathcal{A}}_{ij})$, can be equivalently interpreted as a dynamical metric: 
\ie
\hat{{h}}_{ij} = - \bar{{\ell}}^2 \left(\epsilon_{ik}\hat{\mathcal{A}}_{jk}+ \epsilon_{jk} \hat{\mathcal{A}}_{ik}\right)~, 
\fe
in linearized gravity, which transforms as \eqref{eq:gauge-higher-rank-dual}:
\ie
\hat{{h}}_{ij} \to \hat{{h}}_{ij} - \bar{{\ell}}^2 \left(\epsilon_{ik} \d_j \d_k + \epsilon_{jk} \d_i \d_k \right)\hat \lambda~.
\fe
This establishes an area-preserving diffeomorphism, represented as $x^i \to x^i + \xi^i$, subject to the constraint $\partial_i \xi^i = 0$, with the parameter $\xi^i$ defined as $\xi^i = \bar{{\ell}}^2 \epsilon^{ik} \partial_k \hat \lambda$. The gauge-invariant field strength $\hat{\mathcal{E}}_{ij}$, $\hat{\mathcal{B}}$ in \eqref{eq:gauge-invariant-field-strengths} can be written in terms of this equivalent expression $(\hat{\mathcal{A}}_t, \hat{{h}}_{ij})$ as
\ie
\bar{{\ell}}^2 \left(\epsilon_{ik} \hat{\mathcal{E}}_{jk} + \epsilon_{jk} \hat{\mathcal{E}}_{ik}\right) = \d_i \hat{{v}}_j + \d_j \hat{{v}}_i +\dot{\hat{{h}}}_{ij},~ \bar{{\ell}}^2 \hat{\mathcal{B}} =\frac{1}{2} \hat{\mathcal{R}} =\frac{1}{2} \d_i \d_j \hat{{h}}_{ij}~,
\fe
where the shift vector is expressed as $\hat{{v}}_i= \bar{{\ell}}^2 \epsilon_{ij}\d_j \hat{\mathcal{A}}_t$ and $\hat{\mathcal{R}}$ as the linearized Ricci scalar. Therefore, the Lagrangian for the dual tensor gauge theory \eqref{eq:dual-tensor-theory} can be expressed as a Maxwell-Chern-Simons theory within the framework of linearized gravity.

\bibliography{vortex}

\begin{thebibliography}{50}%
\makeatletter
\providecommand \@ifxundefined [1]{%
 \@ifx{#1\undefined}
}%
\providecommand \@ifnum [1]{%
 \ifnum #1\expandafter \@firstoftwo
 \else \expandafter \@secondoftwo
 \fi
}%
\providecommand \@ifx [1]{%
 \ifx #1\expandafter \@firstoftwo
 \else \expandafter \@secondoftwo
 \fi
}%
\providecommand \natexlab [1]{#1}%
\providecommand \enquote  [1]{``#1''}%
\providecommand \bibnamefont  [1]{#1}%
\providecommand \bibfnamefont [1]{#1}%
\providecommand \citenamefont [1]{#1}%
\providecommand \href@noop [0]{\@secondoftwo}%
\providecommand \href [0]{\begingroup \@sanitize@url \@href}%
\providecommand \@href[1]{\@@startlink{#1}\@@href}%
\providecommand \@@href[1]{\endgroup#1\@@endlink}%
\providecommand \@sanitize@url [0]{\catcode `\\12\catcode `\$12\catcode
  `\&12\catcode `\#12\catcode `\^12\catcode `\_12\catcode `\%12\relax}%
\providecommand \@@startlink[1]{}%
\providecommand \@@endlink[0]{}%
\providecommand \url  [0]{\begingroup\@sanitize@url \@url }%
\providecommand \@url [1]{\endgroup\@href {#1}{\urlprefix }}%
\providecommand \urlprefix  [0]{URL }%
\providecommand \Eprint [0]{\href }%
\providecommand \doibase [0]{https://doi.org/}%
\providecommand \selectlanguage [0]{\@gobble}%
\providecommand \bibinfo  [0]{\@secondoftwo}%
\providecommand \bibfield  [0]{\@secondoftwo}%
\providecommand \translation [1]{[#1]}%
\providecommand \BibitemOpen [0]{}%
\providecommand \bibitemStop [0]{}%
\providecommand \bibitemNoStop [0]{.\EOS\space}%
\providecommand \EOS [0]{\spacefactor3000\relax}%
\providecommand \BibitemShut  [1]{\csname bibitem#1\endcsname}%
\let\auto@bib@innerbib\@empty
\bibitem [{\citenamefont {Haah}(2011)}]{PhysRevA.83.042330}%
  \BibitemOpen
  \bibfield  {author} {\bibinfo {author} {\bibfnamefont {J.}~\bibnamefont
  {Haah}},\ }\bibfield  {title} {\bibinfo {title} {Local stabilizer codes in
  three dimensions without string logical operators},\ }\href
  {https://doi.org/10.1103/PhysRevA.83.042330} {\bibfield  {journal} {\bibinfo
  {journal} {Phys. Rev. A}\ }\textbf {\bibinfo {volume} {83}},\ \bibinfo
  {pages} {042330} (\bibinfo {year} {2011})}\BibitemShut {NoStop}%
\bibitem [{\citenamefont {Brown}\ \emph {et~al.}(2016)\citenamefont {Brown},
  \citenamefont {Loss}, \citenamefont {Pachos}, \citenamefont {Self},\ and\
  \citenamefont {Wootton}}]{RevModPhys.88.045005}%
  \BibitemOpen
  \bibfield  {author} {\bibinfo {author} {\bibfnamefont {B.~J.}\ \bibnamefont
  {Brown}}, \bibinfo {author} {\bibfnamefont {D.}~\bibnamefont {Loss}},
  \bibinfo {author} {\bibfnamefont {J.~K.}\ \bibnamefont {Pachos}}, \bibinfo
  {author} {\bibfnamefont {C.~N.}\ \bibnamefont {Self}},\ and\ \bibinfo
  {author} {\bibfnamefont {J.~R.}\ \bibnamefont {Wootton}},\ }\bibfield
  {title} {\bibinfo {title} {Quantum memories at finite temperature},\ }\href
  {https://doi.org/10.1103/RevModPhys.88.045005} {\bibfield  {journal}
  {\bibinfo  {journal} {Rev. Mod. Phys.}\ }\textbf {\bibinfo {volume} {88}},\
  \bibinfo {pages} {045005} (\bibinfo {year} {2016})}\BibitemShut {NoStop}%
\bibitem [{\citenamefont {Roberts}\ and\ \citenamefont
  {Bartlett}(2020)}]{PhysRevX.10.031041}%
  \BibitemOpen
  \bibfield  {author} {\bibinfo {author} {\bibfnamefont {S.}~\bibnamefont
  {Roberts}}\ and\ \bibinfo {author} {\bibfnamefont {S.~D.}\ \bibnamefont
  {Bartlett}},\ }\bibfield  {title} {\bibinfo {title} {Symmetry-protected
  self-correcting quantum memories},\ }\href
  {https://doi.org/10.1103/PhysRevX.10.031041} {\bibfield  {journal} {\bibinfo
  {journal} {Phys. Rev. X}\ }\textbf {\bibinfo {volume} {10}},\ \bibinfo
  {pages} {031041} (\bibinfo {year} {2020})}\BibitemShut {NoStop}%
\bibitem [{\citenamefont {Beekman}\ \emph {et~al.}(2017)\citenamefont
  {Beekman}, \citenamefont {Nissinen}, \citenamefont {Wu}, \citenamefont {Liu},
  \citenamefont {Slager}, \citenamefont {Nussinov}, \citenamefont {Cvetkovic},\
  and\ \citenamefont {Zaanen}}]{Beekman_2017}%
  \BibitemOpen
  \bibfield  {author} {\bibinfo {author} {\bibfnamefont {A.~J.}\ \bibnamefont
  {Beekman}}, \bibinfo {author} {\bibfnamefont {J.}~\bibnamefont {Nissinen}},
  \bibinfo {author} {\bibfnamefont {K.}~\bibnamefont {Wu}}, \bibinfo {author}
  {\bibfnamefont {K.}~\bibnamefont {Liu}}, \bibinfo {author} {\bibfnamefont
  {R.-J.}\ \bibnamefont {Slager}}, \bibinfo {author} {\bibfnamefont
  {Z.}~\bibnamefont {Nussinov}}, \bibinfo {author} {\bibfnamefont
  {V.}~\bibnamefont {Cvetkovic}},\ and\ \bibinfo {author} {\bibfnamefont
  {J.}~\bibnamefont {Zaanen}},\ }\bibfield  {title} {\bibinfo {title} {Dual
  gauge field theory of quantum liquid crystals in two dimensions},\ }\href
  {https://doi.org/10.1016/j.physrep.2017.03.004} {\bibfield  {journal}
  {\bibinfo  {journal} {Physics Reports}\ }\textbf {\bibinfo {volume} {683}},\
  \bibinfo {pages} {1} (\bibinfo {year} {2017})}\BibitemShut {NoStop}%
\bibitem [{\citenamefont {Pretko}\ and\ \citenamefont
  {Radzihovsky}(2018{\natexlab{a}})}]{Pretko_2018}%
  \BibitemOpen
  \bibfield  {author} {\bibinfo {author} {\bibfnamefont {M.}~\bibnamefont
  {Pretko}}\ and\ \bibinfo {author} {\bibfnamefont {L.}~\bibnamefont
  {Radzihovsky}},\ }\bibfield  {title} {\bibinfo {title} {Fracton-elasticity
  duality},\ }\bibfield  {journal} {\bibinfo  {journal} {Physical Review
  Letters}\ }\textbf {\bibinfo {volume} {120}},\ \href
  {https://doi.org/10.1103/physrevlett.120.195301}
  {10.1103/physrevlett.120.195301} (\bibinfo {year}
  {2018}{\natexlab{a}})\BibitemShut {NoStop}%
\bibitem [{\citenamefont {Pretko}\ and\ \citenamefont
  {Radzihovsky}(2018{\natexlab{b}})}]{PhysRevLett.121.235301}%
  \BibitemOpen
  \bibfield  {author} {\bibinfo {author} {\bibfnamefont {M.}~\bibnamefont
  {Pretko}}\ and\ \bibinfo {author} {\bibfnamefont {L.}~\bibnamefont
  {Radzihovsky}},\ }\bibfield  {title} {\bibinfo {title} {Symmetry-enriched
  fracton phases from supersolid duality},\ }\href
  {https://doi.org/10.1103/PhysRevLett.121.235301} {\bibfield  {journal}
  {\bibinfo  {journal} {Phys. Rev. Lett.}\ }\textbf {\bibinfo {volume} {121}},\
  \bibinfo {pages} {235301} (\bibinfo {year} {2018}{\natexlab{b}})}\BibitemShut
  {NoStop}%
\bibitem [{\citenamefont {Gaiotto}\ \emph {et~al.}(2015)\citenamefont
  {Gaiotto}, \citenamefont {Kapustin}, \citenamefont {Seiberg},\ and\
  \citenamefont {Willett}}]{Gaiotto_2015}%
  \BibitemOpen
  \bibfield  {author} {\bibinfo {author} {\bibfnamefont {D.}~\bibnamefont
  {Gaiotto}}, \bibinfo {author} {\bibfnamefont {A.}~\bibnamefont {Kapustin}},
  \bibinfo {author} {\bibfnamefont {N.}~\bibnamefont {Seiberg}},\ and\ \bibinfo
  {author} {\bibfnamefont {B.}~\bibnamefont {Willett}},\ }\bibfield  {title}
  {\bibinfo {title} {Generalized global symmetries},\ }\bibfield  {journal}
  {\bibinfo  {journal} {Journal of High Energy Physics}\ }\textbf {\bibinfo
  {volume} {2015}},\ \href {https://doi.org/10.1007/jhep02(2015)172}
  {10.1007/jhep02(2015)172} (\bibinfo {year} {2015})\BibitemShut {NoStop}%
\bibitem [{\citenamefont {Qi}\ \emph {et~al.}(2021)\citenamefont {Qi},
  \citenamefont {Radzihovsky},\ and\ \citenamefont {Hermele}}]{Qi_2021}%
  \BibitemOpen
  \bibfield  {author} {\bibinfo {author} {\bibfnamefont {M.}~\bibnamefont
  {Qi}}, \bibinfo {author} {\bibfnamefont {L.}~\bibnamefont {Radzihovsky}},\
  and\ \bibinfo {author} {\bibfnamefont {M.}~\bibnamefont {Hermele}},\
  }\bibfield  {title} {\bibinfo {title} {Fracton phases via exotic higher-form
  symmetry-breaking},\ }\href {https://doi.org/10.1016/j.aop.2020.168360}
  {\bibfield  {journal} {\bibinfo  {journal} {Annals of Physics}\ }\textbf
  {\bibinfo {volume} {424}},\ \bibinfo {pages} {168360} (\bibinfo {year}
  {2021})}\BibitemShut {NoStop}%
\bibitem [{\citenamefont {Nandkishore}\ and\ \citenamefont
  {Hermele}(2019)}]{Nandkishore_2019}%
  \BibitemOpen
  \bibfield  {author} {\bibinfo {author} {\bibfnamefont {R.~M.}\ \bibnamefont
  {Nandkishore}}\ and\ \bibinfo {author} {\bibfnamefont {M.}~\bibnamefont
  {Hermele}},\ }\bibfield  {title} {\bibinfo {title} {Fractons},\ }\href
  {https://doi.org/10.1146/annurev-conmatphys-031218-013604} {\bibfield
  {journal} {\bibinfo  {journal} {Annual Review of Condensed Matter Physics}\
  }\textbf {\bibinfo {volume} {10}},\ \bibinfo {pages} {295} (\bibinfo {year}
  {2019})}\BibitemShut {NoStop}%
\bibitem [{\citenamefont {Pretko}\ \emph {et~al.}(2020)\citenamefont {Pretko},
  \citenamefont {Chen},\ and\ \citenamefont {You}}]{Pretko_2020}%
  \BibitemOpen
  \bibfield  {author} {\bibinfo {author} {\bibfnamefont {M.}~\bibnamefont
  {Pretko}}, \bibinfo {author} {\bibfnamefont {X.}~\bibnamefont {Chen}},\ and\
  \bibinfo {author} {\bibfnamefont {Y.}~\bibnamefont {You}},\ }\bibfield
  {title} {\bibinfo {title} {Fracton phases of matter},\ }\href
  {https://doi.org/10.1142/s0217751x20300033} {\bibfield  {journal} {\bibinfo
  {journal} {International Journal of Modern Physics A}\ }\textbf {\bibinfo
  {volume} {35}},\ \bibinfo {pages} {2030003} (\bibinfo {year}
  {2020})}\BibitemShut {NoStop}%
\bibitem [{\citenamefont {Du}\ \emph {et~al.}(2022)\citenamefont {Du},
  \citenamefont {Mehta}, \citenamefont {Nguyen},\ and\ \citenamefont
  {Son}}]{Du_2022}%
  \BibitemOpen
  \bibfield  {author} {\bibinfo {author} {\bibfnamefont {Y.-H.}\ \bibnamefont
  {Du}}, \bibinfo {author} {\bibfnamefont {U.}~\bibnamefont {Mehta}}, \bibinfo
  {author} {\bibfnamefont {D.}~\bibnamefont {Nguyen}},\ and\ \bibinfo {author}
  {\bibfnamefont {D.~T.}\ \bibnamefont {Son}},\ }\bibfield  {title} {\bibinfo
  {title} {Volume-preserving diffeomorphism as nonabelian higher-rank gauge
  symmetry},\ }\bibfield  {journal} {\bibinfo  {journal} {{SciPost} Physics}\
  }\textbf {\bibinfo {volume} {12}},\ \href
  {https://doi.org/10.21468/scipostphys.12.2.050}
  {10.21468/scipostphys.12.2.050} (\bibinfo {year} {2022})\BibitemShut
  {NoStop}%
\bibitem [{\citenamefont {Gromov}\ and\ \citenamefont
  {Radzihovsky}(2024)}]{RevModPhys.96.011001}%
  \BibitemOpen
  \bibfield  {author} {\bibinfo {author} {\bibfnamefont {A.}~\bibnamefont
  {Gromov}}\ and\ \bibinfo {author} {\bibfnamefont {L.}~\bibnamefont
  {Radzihovsky}},\ }\bibfield  {title} {\bibinfo {title} {Colloquium: Fracton
  matter},\ }\href {https://doi.org/10.1103/RevModPhys.96.011001} {\bibfield
  {journal} {\bibinfo  {journal} {Rev. Mod. Phys.}\ }\textbf {\bibinfo {volume}
  {96}},\ \bibinfo {pages} {011001} (\bibinfo {year} {2024})}\BibitemShut
  {NoStop}%
\bibitem [{\citenamefont {Du}\ \emph {et~al.}(2024{\natexlab{a}})\citenamefont
  {Du}, \citenamefont {Moroz}, \citenamefont {Nguyen},\ and\ \citenamefont
  {Son}}]{PhysRevResearch.6.L012040}%
  \BibitemOpen
  \bibfield  {author} {\bibinfo {author} {\bibfnamefont {Y.-H.}\ \bibnamefont
  {Du}}, \bibinfo {author} {\bibfnamefont {S.}~\bibnamefont {Moroz}}, \bibinfo
  {author} {\bibfnamefont {D.~X.}\ \bibnamefont {Nguyen}},\ and\ \bibinfo
  {author} {\bibfnamefont {D.~T.}\ \bibnamefont {Son}},\ }\bibfield  {title}
  {\bibinfo {title} {Noncommutative field theory of the tkachenko mode:
  Symmetries and decay rate},\ }\href
  {https://doi.org/10.1103/PhysRevResearch.6.L012040} {\bibfield  {journal}
  {\bibinfo  {journal} {Phys. Rev. Res.}\ }\textbf {\bibinfo {volume} {6}},\
  \bibinfo {pages} {L012040} (\bibinfo {year}
  {2024}{\natexlab{a}})}\BibitemShut {NoStop}%
\bibitem [{\citenamefont {Pretko}\ \emph {et~al.}(2019)\citenamefont {Pretko},
  \citenamefont {Zhai},\ and\ \citenamefont
  {Radzihovsky}}]{PhysRevB.100.134113}%
  \BibitemOpen
  \bibfield  {author} {\bibinfo {author} {\bibfnamefont {M.}~\bibnamefont
  {Pretko}}, \bibinfo {author} {\bibfnamefont {Z.}~\bibnamefont {Zhai}},\ and\
  \bibinfo {author} {\bibfnamefont {L.}~\bibnamefont {Radzihovsky}},\
  }\bibfield  {title} {\bibinfo {title} {Crystal-to-fracton tensor gauge theory
  dualities},\ }\href {https://doi.org/10.1103/PhysRevB.100.134113} {\bibfield
  {journal} {\bibinfo  {journal} {Phys. Rev. B}\ }\textbf {\bibinfo {volume}
  {100}},\ \bibinfo {pages} {134113} (\bibinfo {year} {2019})}\BibitemShut
  {NoStop}%
\bibitem [{\citenamefont {Radzihovsky}\ and\ \citenamefont
  {Hermele}(2020)}]{RHvectorPRL2020}%
  \BibitemOpen
  \bibfield  {author} {\bibinfo {author} {\bibfnamefont {L.}~\bibnamefont
  {Radzihovsky}}\ and\ \bibinfo {author} {\bibfnamefont {M.}~\bibnamefont
  {Hermele}},\ }\bibfield  {title} {\bibinfo {title} {Fractons from vector
  gauge theory},\ }\bibfield  {journal} {\bibinfo  {journal} {Physical Review
  Letters}\ }\textbf {\bibinfo {volume} {124}},\ \href
  {https://doi.org/10.1103/physrevlett.124.050402}
  {10.1103/physrevlett.124.050402} (\bibinfo {year} {2020})\BibitemShut
  {NoStop}%
\bibitem [{\citenamefont {Radzihovsky}(2020)}]{PhysRevLett.125.267601}%
  \BibitemOpen
  \bibfield  {author} {\bibinfo {author} {\bibfnamefont {L.}~\bibnamefont
  {Radzihovsky}},\ }\bibfield  {title} {\bibinfo {title} {Quantum smectic gauge
  theory},\ }\href {https://doi.org/10.1103/PhysRevLett.125.267601} {\bibfield
  {journal} {\bibinfo  {journal} {Phys. Rev. Lett.}\ }\textbf {\bibinfo
  {volume} {125}},\ \bibinfo {pages} {267601} (\bibinfo {year}
  {2020})}\BibitemShut {NoStop}%
\bibitem [{\citenamefont {Zhai}\ and\ \citenamefont
  {Radzihovsky}(2021)}]{Zhai_2021}%
  \BibitemOpen
  \bibfield  {author} {\bibinfo {author} {\bibfnamefont {Z.}~\bibnamefont
  {Zhai}}\ and\ \bibinfo {author} {\bibfnamefont {L.}~\bibnamefont
  {Radzihovsky}},\ }\bibfield  {title} {\bibinfo {title} {Fractonic gauge
  theory of smectics},\ }\href {https://doi.org/10.1016/j.aop.2021.168509}
  {\bibfield  {journal} {\bibinfo  {journal} {Annals of Physics}\ }\textbf
  {\bibinfo {volume} {435}},\ \bibinfo {pages} {168509} (\bibinfo {year}
  {2021})}\BibitemShut {NoStop}%
\bibitem [{\citenamefont {Kumar}\ and\ \citenamefont
  {Potter}(2019)}]{Kumar_2019}%
  \BibitemOpen
  \bibfield  {author} {\bibinfo {author} {\bibfnamefont {A.}~\bibnamefont
  {Kumar}}\ and\ \bibinfo {author} {\bibfnamefont {A.~C.}\ \bibnamefont
  {Potter}},\ }\bibfield  {title} {\bibinfo {title} {Symmetry-enforced
  fractonicity and two-dimensional quantum crystal melting},\ }\bibfield
  {journal} {\bibinfo  {journal} {Physical Review B}\ }\textbf {\bibinfo
  {volume} {100}},\ \href {https://doi.org/10.1103/physrevb.100.045119}
  {10.1103/physrevb.100.045119} (\bibinfo {year} {2019})\BibitemShut {NoStop}%
\bibitem [{\citenamefont {Nguyen}\ \emph {et~al.}(2020)\citenamefont {Nguyen},
  \citenamefont {Gromov},\ and\ \citenamefont {Moroz}}]{Nguyen-Gromov-Moroz}%
  \BibitemOpen
  \bibfield  {author} {\bibinfo {author} {\bibfnamefont {D.~X.}\ \bibnamefont
  {Nguyen}}, \bibinfo {author} {\bibfnamefont {A.}~\bibnamefont {Gromov}},\
  and\ \bibinfo {author} {\bibfnamefont {S.}~\bibnamefont {Moroz}},\ }\bibfield
   {title} {\bibinfo {title} {{Fracton-elasticity duality of two-dimensional
  superfluid vortex crystals: defect interactions and quantum melting}},\
  }\href {https://doi.org/10.21468/SciPostPhys.9.5.076} {\bibfield  {journal}
  {\bibinfo  {journal} {SciPost Phys.}\ }\textbf {\bibinfo {volume} {9}},\
  \bibinfo {pages} {076} (\bibinfo {year} {2020})}\BibitemShut {NoStop}%
\bibitem [{\citenamefont {Frey}\ \emph {et~al.}(1994)\citenamefont {Frey},
  \citenamefont {Nelson},\ and\ \citenamefont {Fisher}}]{FreyFisherNelsonSS}%
  \BibitemOpen
  \bibfield  {author} {\bibinfo {author} {\bibfnamefont {E.}~\bibnamefont
  {Frey}}, \bibinfo {author} {\bibfnamefont {D.~R.}\ \bibnamefont {Nelson}},\
  and\ \bibinfo {author} {\bibfnamefont {D.~S.}\ \bibnamefont {Fisher}},\
  }\bibfield  {title} {\bibinfo {title} {Interstitials, vacancies, and
  supersolid order in vortex crystals},\ }\href
  {https://doi.org/10.1103/PhysRevB.49.9723} {\bibfield  {journal} {\bibinfo
  {journal} {Phys. Rev. B}\ }\textbf {\bibinfo {volume} {49}},\ \bibinfo
  {pages} {9723} (\bibinfo {year} {1994})}\BibitemShut {NoStop}%
\bibitem [{\citenamefont {Marchetti}\ and\ \citenamefont
  {Radzihovsky}(1999)}]{MarchettiLRSS}%
  \BibitemOpen
  \bibfield  {author} {\bibinfo {author} {\bibfnamefont {M.~C.}\ \bibnamefont
  {Marchetti}}\ and\ \bibinfo {author} {\bibfnamefont {L.}~\bibnamefont
  {Radzihovsky}},\ }\bibfield  {title} {\bibinfo {title} {Interstitials,
  vacancies, and dislocations in flux-line lattices: A theory of vortex
  crystals, supersolids, and liquids},\ }\href
  {https://doi.org/10.1103/physrevb.59.12001} {\bibfield  {journal} {\bibinfo
  {journal} {Physical Review B}\ }\textbf {\bibinfo {volume} {59}},\ \bibinfo
  {pages} {12001} (\bibinfo {year} {1999})}\BibitemShut {NoStop}%
\bibitem [{\citenamefont {Moessner}\ \emph {et~al.}(2001)\citenamefont
  {Moessner}, \citenamefont {Sondhi},\ and\ \citenamefont
  {Fradkin}}]{PhysRevB.65.024504}%
  \BibitemOpen
  \bibfield  {author} {\bibinfo {author} {\bibfnamefont {R.}~\bibnamefont
  {Moessner}}, \bibinfo {author} {\bibfnamefont {S.~L.}\ \bibnamefont
  {Sondhi}},\ and\ \bibinfo {author} {\bibfnamefont {E.}~\bibnamefont
  {Fradkin}},\ }\bibfield  {title} {\bibinfo {title} {Short-ranged resonating
  valence bond physics, quantum dimer models, and ising gauge theories},\
  }\href {https://doi.org/10.1103/PhysRevB.65.024504} {\bibfield  {journal}
  {\bibinfo  {journal} {Phys. Rev. B}\ }\textbf {\bibinfo {volume} {65}},\
  \bibinfo {pages} {024504} (\bibinfo {year} {2001})}\BibitemShut {NoStop}%
\bibitem [{\citenamefont {Vishwanath}\ \emph {et~al.}(2004)\citenamefont
  {Vishwanath}, \citenamefont {Balents},\ and\ \citenamefont
  {Senthil}}]{PhysRevB.69.224416}%
  \BibitemOpen
  \bibfield  {author} {\bibinfo {author} {\bibfnamefont {A.}~\bibnamefont
  {Vishwanath}}, \bibinfo {author} {\bibfnamefont {L.}~\bibnamefont
  {Balents}},\ and\ \bibinfo {author} {\bibfnamefont {T.}~\bibnamefont
  {Senthil}},\ }\bibfield  {title} {\bibinfo {title} {Quantum criticality and
  deconfinement in phase transitions between valence bond solids},\ }\href
  {https://doi.org/10.1103/PhysRevB.69.224416} {\bibfield  {journal} {\bibinfo
  {journal} {Phys. Rev. B}\ }\textbf {\bibinfo {volume} {69}},\ \bibinfo
  {pages} {224416} (\bibinfo {year} {2004})}\BibitemShut {NoStop}%
\bibitem [{\citenamefont {Fradkin}\ \emph {et~al.}(2004)\citenamefont
  {Fradkin}, \citenamefont {Huse}, \citenamefont {Moessner}, \citenamefont
  {Oganesyan},\ and\ \citenamefont {Sondhi}}]{PhysRevB.69.224415}%
  \BibitemOpen
  \bibfield  {author} {\bibinfo {author} {\bibfnamefont {E.}~\bibnamefont
  {Fradkin}}, \bibinfo {author} {\bibfnamefont {D.~A.}\ \bibnamefont {Huse}},
  \bibinfo {author} {\bibfnamefont {R.}~\bibnamefont {Moessner}}, \bibinfo
  {author} {\bibfnamefont {V.}~\bibnamefont {Oganesyan}},\ and\ \bibinfo
  {author} {\bibfnamefont {S.~L.}\ \bibnamefont {Sondhi}},\ }\bibfield  {title}
  {\bibinfo {title} {Bipartite rokhsar--kivelson points and cantor
  deconfinement},\ }\href {https://doi.org/10.1103/PhysRevB.69.224415}
  {\bibfield  {journal} {\bibinfo  {journal} {Phys. Rev. B}\ }\textbf {\bibinfo
  {volume} {69}},\ \bibinfo {pages} {224415} (\bibinfo {year}
  {2004})}\BibitemShut {NoStop}%
\bibitem [{\citenamefont {Ardonne}\ \emph {et~al.}(2004)\citenamefont
  {Ardonne}, \citenamefont {Fendley},\ and\ \citenamefont
  {Fradkin}}]{Ardonne_2004}%
  \BibitemOpen
  \bibfield  {author} {\bibinfo {author} {\bibfnamefont {E.}~\bibnamefont
  {Ardonne}}, \bibinfo {author} {\bibfnamefont {P.}~\bibnamefont {Fendley}},\
  and\ \bibinfo {author} {\bibfnamefont {E.}~\bibnamefont {Fradkin}},\
  }\bibfield  {title} {\bibinfo {title} {Topological order and conformal
  quantum critical points},\ }\href {https://doi.org/10.1016/j.aop.2004.01.004}
  {\bibfield  {journal} {\bibinfo  {journal} {Annals of Physics}\ }\textbf
  {\bibinfo {volume} {310}},\ \bibinfo {pages} {493} (\bibinfo {year}
  {2004})}\BibitemShut {NoStop}%
\bibitem [{\citenamefont {Gorantla}\ \emph {et~al.}(2022)\citenamefont
  {Gorantla}, \citenamefont {Lam}, \citenamefont {Seiberg},\ and\ \citenamefont
  {Shao}}]{Gorantla:2022eem}%
  \BibitemOpen
  \bibfield  {author} {\bibinfo {author} {\bibfnamefont {P.}~\bibnamefont
  {Gorantla}}, \bibinfo {author} {\bibfnamefont {H.~T.}\ \bibnamefont {Lam}},
  \bibinfo {author} {\bibfnamefont {N.}~\bibnamefont {Seiberg}},\ and\ \bibinfo
  {author} {\bibfnamefont {S.-H.}\ \bibnamefont {Shao}},\ }\bibfield  {title}
  {\bibinfo {title} {{Global dipole symmetry, compact Lifshitz theory, tensor
  gauge theory, and fractons}},\ }\href
  {https://doi.org/10.1103/PhysRevB.106.045112} {\bibfield  {journal} {\bibinfo
   {journal} {Phys. Rev. B}\ }\textbf {\bibinfo {volume} {106}},\ \bibinfo
  {pages} {045112} (\bibinfo {year} {2022})},\ \Eprint
  {https://arxiv.org/abs/2201.10589} {arXiv:2201.10589 [cond-mat.str-el]}
  \BibitemShut {NoStop}%
\bibitem [{\citenamefont {Lake}\ \emph {et~al.}(2022)\citenamefont {Lake},
  \citenamefont {Hermele},\ and\ \citenamefont
  {Senthil}}]{PhysRevB.106.064511}%
  \BibitemOpen
  \bibfield  {author} {\bibinfo {author} {\bibfnamefont {E.}~\bibnamefont
  {Lake}}, \bibinfo {author} {\bibfnamefont {M.}~\bibnamefont {Hermele}},\ and\
  \bibinfo {author} {\bibfnamefont {T.}~\bibnamefont {Senthil}},\ }\bibfield
  {title} {\bibinfo {title} {Dipolar bose-hubbard model},\ }\href
  {https://doi.org/10.1103/PhysRevB.106.064511} {\bibfield  {journal} {\bibinfo
   {journal} {Phys. Rev. B}\ }\textbf {\bibinfo {volume} {106}},\ \bibinfo
  {pages} {064511} (\bibinfo {year} {2022})}\BibitemShut {NoStop}%
\bibitem [{\citenamefont {Gorantla}\ \emph {et~al.}(2023)\citenamefont
  {Gorantla}, \citenamefont {Lam}, \citenamefont {Seiberg},\ and\ \citenamefont
  {Shao}}]{Gorantla:2022ssr}%
  \BibitemOpen
  \bibfield  {author} {\bibinfo {author} {\bibfnamefont {P.}~\bibnamefont
  {Gorantla}}, \bibinfo {author} {\bibfnamefont {H.~T.}\ \bibnamefont {Lam}},
  \bibinfo {author} {\bibfnamefont {N.}~\bibnamefont {Seiberg}},\ and\ \bibinfo
  {author} {\bibfnamefont {S.-H.}\ \bibnamefont {Shao}},\ }\bibfield  {title}
  {\bibinfo {title} {{(2+1)-dimensional compact Lifshitz theory, tensor gauge
  theory, and fractons}},\ }\href {https://doi.org/10.1103/PhysRevB.108.075106}
  {\bibfield  {journal} {\bibinfo  {journal} {Phys. Rev. B}\ }\textbf {\bibinfo
  {volume} {108}},\ \bibinfo {pages} {075106} (\bibinfo {year} {2023})},\
  \Eprint {https://arxiv.org/abs/2209.10030} {arXiv:2209.10030
  [cond-mat.str-el]} \BibitemShut {NoStop}%
\bibitem [{\citenamefont {Radzihovsky}(2022)}]{Radzihovsky_2022}%
  \BibitemOpen
  \bibfield  {author} {\bibinfo {author} {\bibfnamefont {L.}~\bibnamefont
  {Radzihovsky}},\ }\bibfield  {title} {\bibinfo {title} {Lifshitz gauge
  duality},\ }\bibfield  {journal} {\bibinfo  {journal} {Physical Review B}\
  }\textbf {\bibinfo {volume} {106}},\ \href
  {https://doi.org/10.1103/physrevb.106.224510} {10.1103/physrevb.106.224510}
  (\bibinfo {year} {2022})\BibitemShut {NoStop}%
\bibitem [{\citenamefont {Peskin}(1978)}]{peskin1978}%
  \BibitemOpen
  \bibfield  {author} {\bibinfo {author} {\bibfnamefont {M.~E.}\ \bibnamefont
  {Peskin}},\ }\bibfield  {title} {\bibinfo {title} {{Mandelstam-'t Hooft
  Duality in Abelian Lattice Models}},\ }\href
  {https://doi.org/10.1016/0003-4916(78)90252-X} {\bibfield  {journal}
  {\bibinfo  {journal} {Ann. Phys.}\ }\textbf {\bibinfo {volume} {113}},\
  \bibinfo {pages} {122} (\bibinfo {year} {1978})}\BibitemShut {NoStop}%
\bibitem [{\citenamefont {Dasgupta}\ and\ \citenamefont
  {Halperin}(1981)}]{DasguptaHalperin}%
  \BibitemOpen
  \bibfield  {author} {\bibinfo {author} {\bibfnamefont {C.}~\bibnamefont
  {Dasgupta}}\ and\ \bibinfo {author} {\bibfnamefont {B.~I.}\ \bibnamefont
  {Halperin}},\ }\bibfield  {title} {\bibinfo {title} {{Phase Transition in a
  Lattice Model of Superconductivity}},\ }\href
  {https://doi.org/10.1103/PhysRevLett.47.1556} {\bibfield  {journal} {\bibinfo
   {journal} {Phys. Rev. Lett.}\ }\textbf {\bibinfo {volume} {47}},\ \bibinfo
  {pages} {1556} (\bibinfo {year} {1981})}\BibitemShut {NoStop}%
\bibitem [{\citenamefont {Fisher}\ and\ \citenamefont
  {Lee}(1989)}]{PhysRevB.39.2756}%
  \BibitemOpen
  \bibfield  {author} {\bibinfo {author} {\bibfnamefont {M.~P.~A.}\
  \bibnamefont {Fisher}}\ and\ \bibinfo {author} {\bibfnamefont {D.~H.}\
  \bibnamefont {Lee}},\ }\bibfield  {title} {\bibinfo {title} {Correspondence
  between two-dimensional bosons and a bulk superconductor in a magnetic
  field},\ }\href {https://doi.org/10.1103/PhysRevB.39.2756} {\bibfield
  {journal} {\bibinfo  {journal} {Phys. Rev. B}\ }\textbf {\bibinfo {volume}
  {39}},\ \bibinfo {pages} {2756} (\bibinfo {year} {1989})}\BibitemShut
  {NoStop}%
\bibitem [{\citenamefont {Tkachenko}(1965)}]{tkachenko1965}%
  \BibitemOpen
  \bibfield  {author} {\bibinfo {author} {\bibfnamefont {V.~K.}\ \bibnamefont
  {Tkachenko}},\ }\bibfield  {title} {\bibinfo {title} {On vortex lattices},\
  }\href {http://www.jetp.ras.ru/cgi-bin/e/index/e/22/6/p1282?a=list}
  {\bibfield  {journal} {\bibinfo  {journal} {Sov. Phys. JETP}\ }\textbf
  {\bibinfo {volume} {22}},\ \bibinfo {pages} {1282} (\bibinfo {year}
  {1965})}\BibitemShut {NoStop}%
\bibitem [{\citenamefont {Tkachenko}(1966)}]{tkachenko1966}%
  \BibitemOpen
  \bibfield  {author} {\bibinfo {author} {\bibfnamefont {V.~K.}\ \bibnamefont
  {Tkachenko}},\ }\bibfield  {title} {\bibinfo {title} {Stability of vortex
  lattices},\ }\href
  {http://www.jetp.ras.ru/cgi-bin/e/index/e/23/6/p1049?a=list} {\bibfield
  {journal} {\bibinfo  {journal} {Sov. Phys. JETP}\ }\textbf {\bibinfo {volume}
  {23}},\ \bibinfo {pages} {1049} (\bibinfo {year} {1966})}\BibitemShut
  {NoStop}%
\bibitem [{\citenamefont {Tkachenko}(1969)}]{tkachenko1969}%
  \BibitemOpen
  \bibfield  {author} {\bibinfo {author} {\bibfnamefont {V.~K.}\ \bibnamefont
  {Tkachenko}},\ }\bibfield  {title} {\bibinfo {title} {Elasticity of vortex
  lattices},\ }\href
  {http://www.jetp.ras.ru/cgi-bin/e/index/e/29/5/p945?a=list} {\bibfield
  {journal} {\bibinfo  {journal} {Sov. Phys. JETP}\ }\textbf {\bibinfo {volume}
  {29}},\ \bibinfo {pages} {945} (\bibinfo {year} {1969})}\BibitemShut
  {NoStop}%
\bibitem [{\citenamefont {Sonin}(2014)}]{Sonin_2014}%
  \BibitemOpen
  \bibfield  {author} {\bibinfo {author} {\bibfnamefont {E.~B.}\ \bibnamefont
  {Sonin}},\ }\bibfield  {title} {\bibinfo {title} {Tkachenko waves},\ }\href
  {https://doi.org/10.1134/s0021364013240181} {\bibfield  {journal} {\bibinfo
  {journal} {{JETP} Letters}\ }\textbf {\bibinfo {volume} {98}},\ \bibinfo
  {pages} {758} (\bibinfo {year} {2014})}\BibitemShut {NoStop}%
\bibitem [{\citenamefont {Du}\ \emph {et~al.}(2024{\natexlab{b}})\citenamefont
  {Du}, \citenamefont {Xu},\ and\ \citenamefont {Son}}]{PhysRevB.109.035135}%
  \BibitemOpen
  \bibfield  {author} {\bibinfo {author} {\bibfnamefont {Y.-H.}\ \bibnamefont
  {Du}}, \bibinfo {author} {\bibfnamefont {C.}~\bibnamefont {Xu}},\ and\
  \bibinfo {author} {\bibfnamefont {D.~T.}\ \bibnamefont {Son}},\ }\bibfield
  {title} {\bibinfo {title} {Nonlinear lifshitz photon theory in condensed
  matter systems},\ }\href {https://doi.org/10.1103/PhysRevB.109.035135}
  {\bibfield  {journal} {\bibinfo  {journal} {Phys. Rev. B}\ }\textbf {\bibinfo
  {volume} {109}},\ \bibinfo {pages} {035135} (\bibinfo {year}
  {2024}{\natexlab{b}})}\BibitemShut {NoStop}%
\bibitem [{\citenamefont {Douglas}\ and\ \citenamefont
  {Nekrasov}(2001)}]{Douglas:2001ba}%
  \BibitemOpen
  \bibfield  {author} {\bibinfo {author} {\bibfnamefont {M.~R.}\ \bibnamefont
  {Douglas}}\ and\ \bibinfo {author} {\bibfnamefont {N.~A.}\ \bibnamefont
  {Nekrasov}},\ }\bibfield  {title} {\bibinfo {title} {{Noncommutative field
  theory}},\ }\href {https://doi.org/10.1103/RevModPhys.73.977} {\bibfield
  {journal} {\bibinfo  {journal} {Rev. Mod. Phys.}\ }\textbf {\bibinfo {volume}
  {73}},\ \bibinfo {pages} {977} (\bibinfo {year} {2001})},\ \Eprint
  {https://arxiv.org/abs/hep-th/0106048} {hep-th/0106048} \BibitemShut
  {NoStop}%
\bibitem [{\citenamefont {Rubakov}(2005)}]{Rubakov-NC}%
  \BibitemOpen
  \bibfield  {author} {\bibinfo {author} {\bibfnamefont {V.~A.}\ \bibnamefont
  {Rubakov}},\ }\href@noop {} {\emph {\bibinfo {title} {Classical Field
  Theories. Theories with Fermions. Noncommutative Theories}}}\ (\bibinfo
  {publisher} {URSS},\ \bibinfo {year} {2005})\ \bibinfo {note} {(in
  Russian)}\BibitemShut {NoStop}%
\bibitem [{\citenamefont {Du}\ \emph {et~al.}(2021)\citenamefont {Du},
  \citenamefont {Mehta},\ and\ \citenamefont {Son}}]{du2021noncommutative}%
  \BibitemOpen
  \bibfield  {author} {\bibinfo {author} {\bibfnamefont {Y.-H.}\ \bibnamefont
  {Du}}, \bibinfo {author} {\bibfnamefont {U.}~\bibnamefont {Mehta}},\ and\
  \bibinfo {author} {\bibfnamefont {D.~T.}\ \bibnamefont {Son}},\ }\href@noop
  {} {\bibinfo {title} {{Noncommutative gauge symmetry in the fractional
  quantum Hall effect}}} (\bibinfo {year} {2021}),\ \Eprint
  {https://arxiv.org/abs/2110.13875} {arXiv:2110.13875 [cond-mat.str-el]}
  \BibitemShut {NoStop}%
\bibitem [{\citenamefont {Moroz}\ \emph {et~al.}(2018)\citenamefont {Moroz},
  \citenamefont {Hoyos}, \citenamefont {Benzoni},\ and\ \citenamefont
  {Son}}]{Moroz_2018}%
  \BibitemOpen
  \bibfield  {author} {\bibinfo {author} {\bibfnamefont {S.}~\bibnamefont
  {Moroz}}, \bibinfo {author} {\bibfnamefont {C.}~\bibnamefont {Hoyos}},
  \bibinfo {author} {\bibfnamefont {C.}~\bibnamefont {Benzoni}},\ and\ \bibinfo
  {author} {\bibfnamefont {D.~T.}\ \bibnamefont {Son}},\ }\bibfield  {title}
  {\bibinfo {title} {Effective field theory of a vortex lattice in a bosonic
  superfluid},\ }\bibfield  {journal} {\bibinfo  {journal} {{SciPost} Physics}\
  }\textbf {\bibinfo {volume} {5}},\ \href
  {https://doi.org/10.21468/scipostphys.5.4.039} {10.21468/scipostphys.5.4.039}
  (\bibinfo {year} {2018})\BibitemShut {NoStop}%
\bibitem [{\citenamefont {Moroz}\ and\ \citenamefont {Son}(2019)}]{Moroz_2019}%
  \BibitemOpen
  \bibfield  {author} {\bibinfo {author} {\bibfnamefont {S.}~\bibnamefont
  {Moroz}}\ and\ \bibinfo {author} {\bibfnamefont {D.~T.}\ \bibnamefont
  {Son}},\ }\bibfield  {title} {\bibinfo {title} {Bosonic superfluid on the
  lowest landau level},\ }\bibfield  {journal} {\bibinfo  {journal} {Physical
  Review Letters}\ }\textbf {\bibinfo {volume} {122}},\ \href
  {https://doi.org/10.1103/physrevlett.122.235301}
  {10.1103/physrevlett.122.235301} (\bibinfo {year} {2019})\BibitemShut
  {NoStop}%
\bibitem [{Note1()}]{Note1}%
  \BibitemOpen
  \bibinfo {note} {For an incompressible fluid, the dispersion is linear
  instead of quadratic. It can be understood as follows. For vanishing fluid
  compressibility, $\chi =0$ and the kinetic term $\protect \frac {\chi
  }{2}(\partial _t\phi )^2$ in the effective Lifshitz theory \protect \eqref
  {eq:phi_Lagrangian} vanishes. It is then necessary to retain the subdominant
  term $\protect \frac {\rho }{2}(\partial _t u_i)^2$ in \protect \eqref {Lel},
  which under \protect \eqref {eq:u_to_phi} becomes $\protect \frac {\rho
  }{2B^2}(\partial _t \partial _i \phi )^2$. The dispersion is then modified to
  a linear one.}\BibitemShut {Stop}%
\bibitem [{\citenamefont {Watanabe}\ and\ \citenamefont
  {Murayama}(2013)}]{Watanabe:2013iia}%
  \BibitemOpen
  \bibfield  {author} {\bibinfo {author} {\bibfnamefont {H.}~\bibnamefont
  {Watanabe}}\ and\ \bibinfo {author} {\bibfnamefont {H.}~\bibnamefont
  {Murayama}},\ }\bibfield  {title} {\bibinfo {title} {{Redundancies in
  Nambu-Goldstone Bosons}},\ }\href
  {https://doi.org/10.1103/PhysRevLett.110.181601} {\bibfield  {journal}
  {\bibinfo  {journal} {Phys. Rev. Lett.}\ }\textbf {\bibinfo {volume} {110}},\
  \bibinfo {pages} {181601} (\bibinfo {year} {2013})},\ \Eprint
  {https://arxiv.org/abs/1302.4800} {arXiv:1302.4800 [cond-mat.other]}
  \BibitemShut {NoStop}%
\bibitem [{\citenamefont {Doshi}\ and\ \citenamefont
  {Gromov}(2021)}]{doshi2021vortices}%
  \BibitemOpen
  \bibfield  {author} {\bibinfo {author} {\bibfnamefont {D.}~\bibnamefont
  {Doshi}}\ and\ \bibinfo {author} {\bibfnamefont {A.}~\bibnamefont {Gromov}},\
  }\bibfield  {title} {\bibinfo {title} {Vortices as fractons},\ }\href@noop {}
  {\bibfield  {journal} {\bibinfo  {journal} {Communications Physics}\ }\textbf
  {\bibinfo {volume} {4}},\ \bibinfo {pages} {44} (\bibinfo {year}
  {2021})}\BibitemShut {NoStop}%
\bibitem [{\citenamefont {Kosterlitz}\ and\ \citenamefont
  {Thouless}(1973)}]{KT}%
  \BibitemOpen
  \bibfield  {author} {\bibinfo {author} {\bibfnamefont {J.~M.}\ \bibnamefont
  {Kosterlitz}}\ and\ \bibinfo {author} {\bibfnamefont {D.~J.}\ \bibnamefont
  {Thouless}},\ }\bibfield  {title} {\bibinfo {title} {{Ordering, metastability
  and phase transitions in two-dimensional systems}},\ }\href
  {https://doi.org/10.1088/0022-3719/6/7/010} {\bibfield  {journal} {\bibinfo
  {journal} {J. Phys. C}\ }\textbf {\bibinfo {volume} {6}},\ \bibinfo {pages}
  {1181} (\bibinfo {year} {1973})}\BibitemShut {NoStop}%
\bibitem [{\citenamefont {Halperin}\ and\ \citenamefont
  {Nelson}(1978)}]{HNmelting}%
  \BibitemOpen
  \bibfield  {author} {\bibinfo {author} {\bibfnamefont {B.~I.}\ \bibnamefont
  {Halperin}}\ and\ \bibinfo {author} {\bibfnamefont {D.~R.}\ \bibnamefont
  {Nelson}},\ }\bibfield  {title} {\bibinfo {title} {Theory of two-dimensional
  melting},\ }\href {https://doi.org/10.1103/PhysRevLett.41.121} {\bibfield
  {journal} {\bibinfo  {journal} {Phys. Rev. Lett.}\ }\textbf {\bibinfo
  {volume} {41}},\ \bibinfo {pages} {121} (\bibinfo {year} {1978})}\BibitemShut
  {NoStop}%
\bibitem [{\citenamefont {Young}(1979)}]{Youngmelting}%
  \BibitemOpen
  \bibfield  {author} {\bibinfo {author} {\bibfnamefont {A.~P.}\ \bibnamefont
  {Young}},\ }\bibfield  {title} {\bibinfo {title} {Melting and the vector
  coulomb gas in two dimensions},\ }\href
  {https://doi.org/10.1103/PhysRevB.19.1855} {\bibfield  {journal} {\bibinfo
  {journal} {Phys. Rev. B}\ }\textbf {\bibinfo {volume} {19}},\ \bibinfo
  {pages} {1855} (\bibinfo {year} {1979})}\BibitemShut {NoStop}%
\bibitem [{Note2()}]{Note2}%
  \BibitemOpen
  \bibinfo {note} {For topological crystalline defects, we are referring to
  defects associated with field configurations with non-trivial topology. This
  should not be confused with the symmetry operators/defects that implement the
  crystalline symmetries.}\BibitemShut {Stop}%
\bibitem [{\citenamefont {Nguyen}\ and\ \citenamefont
  {Moroz}(2023)}]{Dung:2023}%
  \BibitemOpen
  \bibfield  {author} {\bibinfo {author} {\bibfnamefont {D.~X.}\ \bibnamefont
  {Nguyen}}\ and\ \bibinfo {author} {\bibfnamefont {S.}~\bibnamefont {Moroz}},\
  }\href@noop {} {\bibinfo {title} {On quantum melting of superfluid vortex
  crystals: from lifshitz scalar to dual gravity}} (\bibinfo {year} {2023}),\
  \Eprint {https://arxiv.org/abs/2310.13741} {arXiv:2310.13741
  [cond-mat.quant-gas]} \BibitemShut {NoStop}%
\end{thebibliography}%

\end{document}